\documentclass[preprint2]{aastex}

\newcommand{\bvj}{B-V}

\newcommand{\by}{b-y}
\newcommand{\bvt}{B_T-V_T}
\newcommand{\jk}{J-K}
\newcommand{\vj}{V-J}

\newcommand{\mbol}{$M_{\rm bol}$}

\newcommand{\msol}{$M_{\odot}$}
\newcommand{\teff}{$T_{eff}$}

\newcommand{\kms}{km\,s$^{-1}$}
\newcommand{\logl}{log\,L/L$_{\odot}$}
\newcommand{\logg}{log\,$g$}
\newcommand{\logt}{log\,T$_{eff}$}

\newcommand{\lxlbol}{log(L$_X$/L$_{bol}$)}
\newcommand{\ergs}{erg s$^{-1}$}
\newcommand{\rasstyc}{RASS-ACT/TRC}

\slugcomment{Accepted for AJ, May 20, 2002}

\shorttitle{Post-T Tauri Stars}
\shortauthors{Mamajek, Meyer, \& Liebert}

\begin{document}



\title{Post-T Tauri Stars in the Nearest OB Association}


\author{Eric E. Mamajek, Michael R. Meyer, \& James Liebert}
\affil{Steward Observatory, 
Dept. of Astronomy, 
University of Arizona,
933 N. Cherry Ave., 
Tucson, AZ 85721}

\email{eem@as.arizona.edu}



\begin{abstract}
We present results of a spectroscopic survey of X-ray- and proper 
motion-selected samples of late-type stars in the Lower Centaurus-Crux (LCC) 
and Upper Centaurus-Lupus (UCL) subgroups of the nearest OB association: 
Scorpius-Centaurus. The primary goals of the survey are to determine the 
star-formation history of the OB subgroups, and to assess the frequency of 
accreting stars in a sample dominated by ``post-T Tauri'' pre-main 
sequence (pre-MS) stars. We investigate two samples: 
(1) proper motion candidates from the ACT and TRC astrometric catalogs 
(Hoogerwerf 2000)
with X-ray counterparts in the {\it ROSAT} All-Sky Survey (RASS) Bright
Source Catalog, and 
(2) G and K-type stars in the {\it Hipparcos} catalog found to be candidate 
members by de Zeeuw et al. (1999). We obtained optical spectra of 130 
candidates with the Siding Springs 2.3-m dual-beam spectrograph. Pre-MS 
stars were identified by (1) strong Li $\lambda$6707 absorption, 
(2) subgiant surface gravities, (3) proper motions consistent with Sco-Cen 
membership, and (4) HR diagram positions consistent with being pre-MS. 
We find 93\% of the RASS-ACT/TRC stars to be probable pre-MS members, 
compared to 73\% of the {\it Hipparcos} candidates. We demonstrate
that measuring the gravity-sensitive band-ratio of \ion{Sr}{2} 
$\lambda$4077 to \ion{Fe}{1} $\lambda$4071 is a valuable means of 
discriminating pre-MS and zero-age main sequence (ZAMS) stars. Using secular 
parallaxes, and {\it Hipparcos}, Tycho-2, and 2MASS photometry, we construct 
an HR diagram.
Depending on the choice of published evolutionary tracks, we find the mean
ages of the pre-MS populations to range between 17-23 Myr for LCC, and 
15-22 Myr for UCL. Taking into account observational errors, it
appears that 95\% of the low-mass star-formation in each subgroup
must have occurred in less than 8 Myr (LCC) and 12 Myr (UCL). Using the 
\citet{Bertelli94} tracks, we find main sequence turn-off ages for {\it 
Hipparcos} B-type members to be 16\,$\pm$\,1\,Myr for LCC and 
17\,$\pm$\,1\,Myr for UCL. Contrary to 
previous findings, it appears that LCC is coeval with, or slightly older 
than, UCL. The secular parallaxes of the Sco-Cen pre-MS stars yield distances 
of 85-215\,pc, with 12 of the LCC members lying within 100\,pc of the Sun. 
Only 1 out of 110 (0.9$^{+2.1}_{-0.8}$\%; 1$\sigma$) pre-MS solar-type stars in the 
sample with ages of 13\,$\pm$\,1\,(s.e.)\,$\pm$\,6 (1$\sigma$) Myr
and masses of 1.3\,$\pm$\,0.2 (1$\sigma$) \msol\, shows both enhanced H$\alpha$ emission 
and a K-band excess indicative of accretion from a truncated 
circumstellar disk: the nearby ($d\,\simeq\,86$\,pc) classical
T Tauri star PDS 66.
\end{abstract}




\section{Introduction}

Post-T Tauri stars (PTTSs) are low-mass, pre-main sequence (pre-MS) stars with 
properties intermediate between T Tauri stars found in molecular clouds 
(both ``classical'' with evidence for accretion from a circumstellar disk
and ``weak-lined'' lacking such evidence; CTTSs \& WTTSs; ages $<$ few Myr) 
and zero-age main sequence stars (ZAMS; ages $>$ 30-100 Myr). Although strict 
observational criteria do not exist for classifying PTTSs as such, a working 
definition is a low-mass star ($<$2\,\msol) that is Li-rich compared to 
stars in ZAMS open clusters such as the $\sim$120 Myr-old Pleiades, and whose 
theoretical H-R position (\logt\, and \logl) is 
above the main sequence \citep{Herbig78,Jensen01}. Since these criteria also apply to 
CTTSs and WTTSs, one could argue that in addition, PTTSs should be located in 
regions devoid of nearby molecular gas or nebulosity. Classifying PTTSs by 
these criteria has complications: (1) few young field 
stars not associated with well-studied molecular clouds currently have accurately 
measured distances (hence known luminosities), 
(2) unresolved binarity can make stars with known distances appear more 
luminous, and thus younger, and (3) there is a dispersion in observed Li 
abundances among stars with the same masses and ages in coeval open clusters. 
Pre-main sequence stars exhibit considerable 
chromospheric (H$\alpha$, Ca H \& K) and coronal X-ray emission. Only a few 
PTTS candidates were known before the {\it Einstein} and {\it ROSAT} X-ray 
missions, and X-ray surveys have become the primary means of identifying 
these pre-MS stars.

Investigations of pre-main sequence evolution 
have been hampered by a lack of large samples of well-characterized PTTSs.
This deficit has impacted studies of pre-MS angular momentum evolution 
\citep[e.g.][]{Rebull01,Bouvier97}, stellar multiplicity \citep[e.g.][]{Kohler00}, 
and circumstellar disk evolution 
\citep[e.g.][]{Spangler01, Haisch01}. The nearest post-T Tauri stars also
provide optimal targets for young exoplanet and brown 
dwarf searches \citep[e.g.][]{Lowrance00}. These objects are much more 
luminous early in their evolution, and the closest targets enable characterization
of the smallest orbital radii. With a 
post-T Tauri population in a nearby OB association, we
can address basic questions such as: 
How long does star-formation persist in a giant molecular cloud?
What is the duration of the accretion phase for young solar-type stars?

Identifying a bona fide PTTS sample can be accomplished by searching for 
low-mass members of nearby fossil OB associations. The Sco-Cen OB complex (Sco OB2) 
is the nearest OB association to the Sun (mean subgroup distances range from 
118-145 pc; \citet{deZeeuw99}, hereafter dZ99), and covers roughly 2000 square
degrees ($\sim$5\%) of the sky. The complex is comprised of three kinematic 
subgroups \citep{Blaauw46} with nuclear ages ranging from 5 to 15 Myr,
a molecular cloud currently undergoing star-formation 
\citep[the $\rho$ Oph complex,][]{Wilking89,Blaauw91,deGeus92}, and perhaps several 
smaller cloud complexes in the vicinity (e.g. the Lupus, Corona Australis, 
Chamaeleon, Musca, and Coalsack clouds). The three subgroups are Upper Scorpius 
(US; age 5-6 Myr), Upper Centaurus-Lupus (UCL; age 14-15 Myr), and Lower 
Centaurus-Crux \citep[LCC; age 11-12 Myr][]{deGeus89}. US has been studied 
extensively in recent years \citep[e.g.][and references therein]{Preibisch99}, 
however UCL and LCC 
have received relatively little attention. 

In this work, we investigate the low-mass ($<$2\,\msol) membership of the two oldest 
Sco-Cen OB subgroups (LCC and UCL) utilizing recently available astrometric 
catalogs ({\it Hipparcos}, Astrographic Catalog-Tycho (ACT), Tycho Reference Catalog
(TRC), and {\it Tycho-2}), the 2 Micron All Sky Survey
(2MASS), and the {\it ROSAT} All-Sky Survey (RASS). We conduct a spectroscopic 
survey of two samples: (1) an X-ray-selected sample of late-type stars from 
the kinematic candidate membership lists of \citet{Hoogerwerf00}, and
(2) the G-K type {\it Hipparcos} members of the OB subgroups from dZ99. In 
\S2, we discuss the procedure for selecting candidate pre-MS stars from both samples, and 
\S3 discusses the observations and assembled database. 
\S4 describes the data analysis and characterization of our stellar sample, and 
\S5 discusses the selection of pre-MS stars, sample contamination, and completeness. 
\S6 describes how we construct an H-R diagram for the subgroups, and 
\S7 presents results regarding the ages of the subgroups, their age spreads,
and the frequency of accretion disks around pre-MS stars. 
\S8 discusses the star-formation history of LCC and UCL, and 
\S9 summarizes the findings of our survey.

\section{Selection of Candidate Pre-MS Stars \label{cand}}

\subsection{The {\it Hipparcos} Sample \label{z99}}

DZ99 lists {\it Hipparcos} Sco-Cen members that were selected using {\it both}
de Bruijne's (1999a) refurbished convergent point method and Hoogerwerf 
\& Aguilar's spaghetti method (1999). Their membership lists contained 31 
G-K stars in UCL and 21 G-K stars in LCC (their Table C1). Most of these 
bright stars have been classified in the Michigan Spectral Survey 
\citep[e.g.][]{Houk75}, however 
SIMBAD\footnote{http://simbad.u-strasbg.fr/Simbad} reveals that most have been 
studied no further. We limit the survey to the 30 G-K candidates with 
Michigan luminosity classes IV or V (see Table \ref{tab_hip}). 
Stars with borderline F/G Michigan types were not observed. HIP 63962 and 73777 
met the criteria, but were not observed. DZ99 estimated the contamination by 
G-K-type interlopers of all luminosity classes to be 32\% for LCC and 24\% for 
UCL. Of these 31 stars, 17 also have RASS-BSC X-ray counterparts within 40\arcsec. 

\placetable{tab_hip}

\subsection{The \rasstyc\, Sample \label{rasstycho}}

To identify lower-mass members of an OB association, one can search
for stars whose proper motions are similar to those of high-mass members.
The high-mass membership and moving group solution for each OB subgroup
were determined by dZ99 and \citet{deBruijne99scocen}. Thousands of faint 
stars in the ACT and TRC astrometric catalogs\footnote{The ACT 
\citep{Urban98ACT} and TRC catalogs \citep{Hog98TRC} were used for target
selection for this project in 1999, however we use the photometry and 
astrometry from the Tycho-2 catalog 
\citep[available in 2000;][]{Hog00TYC2a,Hog00TYC2b} in the data analysis. 
The Tycho-2 catalog was a joint USNO/Copenhagen project, and its data 
supercede the contents of the ACT and TRC catalogs.} were identified by 
\citet{Hoogerwerf00} as candidate low-mass LCC and UCL members.
A high degree of contamination from interlopers is expected due to the 
similarity of the space motions of the subgroups to that of the Local Standard of Rest,
compounded by the low galactic latitude of the subgroups.
The selection of ACT/TRC candidate members is described in detail
in \S4 of \citet{Hoogerwerf00}.

The Hoogerwerf ACT/TRC membership lists for LCC and UCL were slightly 
modified, and filtered, to produce the final target list. First, we requested
from R. Hoogerwerf (personal communication) candidate membership lists with 
different color-magnitude constraints from that described in 
\citet{Hoogerwerf00}. The new color-magnitude selection box is essentially 
a polygon defined by the \cite{Schmidt-Kaler82} empirical zero-age main
sequence ($\bvj$\, vs. M$_V$) at the mean distance for each subgroup (dZ99), 
where we take all 
stars $\Delta$M$_V$\,=\,3 mag above and $\Delta$M$_V$\,=\,1 mag below the ZAMS line. 
Hoogerwerf originally 
selected only those stars within $\Delta$M$_V$\,=\,1.5 mag above the ZAMS, however this 
could inadvertently omit younger members or binaries. The selection box contained 
1353 ACT and TRC stars in LCC, and 1874 stars in UCL. In order to target 
low-mass solar-type stars with G-K spectral types, we retained only those 
stars with Johnson $\bvj$\, $\geq$\, 0.58 mag \citep[the unreddened color of 
G0 dwarfs;][]{Drilling00}. No red $\bvj$~ limit was imposed. 
After the color-magnitude selection, we 
retained only those stars which were identified as kinematic members in 
{\it both} the ACT and TRC astrometric catalogs. This final color-selection 
of the ACT/TRC lists resulted in 785 UCL candidates and 679 
LCC candidates.

In order to further filter the target list, we selected only those 
ACT/TRC candidates which had {\it ROSAT} All-Sky Survey Bright Source 
Catalog (RASS-BSC) X-ray counterparts. \citet{Voges99} cross-referenced the 
RASS-BSC with the Tycho catalog and found that 68\% of the optical-X-ray 
correlations were within 13\arcsec, and 90\% of the correlations were within 
25\arcsec. In plotting a histogram of the separation distance between 
RASS-BSC X-ray sources and ACT/TRC stars, we independently find 40\arcsec\, 
to be an optimal search radius. No constraints on X-ray hardness ratio were 
imposed in the target selection. In order to calculate X-ray luminosities, 
we assume the X-ray energy conversion factor for the {\it ROSAT}~PSPC 
detector from \citet{Fleming95}. The linearity of this X-ray efficiency 
relation spans the temperature range of stellar coronae from inactive 
subdwarf stars to extremely active RS CVns and T Tauri stars. 
Unsurprisingly, the kinematic selection of ACT/TRC stars also selected
many of the same stars as in the {\it Hipparcos} sample
(HIP 57524, 59854, 62445, 65423, 66001, 66941, 67522, 75924, 76472, 77135,
77524, 77656, 80636). These stars are retained in the {\it Hipparcos}
sample (Table \ref{tab_hip}), and omitted from the \rasstyc\, list
(Table \ref{tab_rasstyc}). The final target list of 96 \rasstyc\, stars 
(40 LCC, 56 UCL) is given in Table \ref{tab_rasstyc}.

\placetable{tab_rasstyc}

\section{Observations \label{obs}}

Blue and red optical spectra of the pre-MS candidates were taken 
simultaneously with the Dual-Beam Spectrograph (DBS) on the Siding 
Springs 2.3-m telescope on the nights of 20-24 April 2000. The DBS instrument
is detailed in \citet{Rodgers88}. Using a 2\arcsec\, wide slit, we used 
the B600 l/mm grating in first order on the blue channel, yielding 2.8\AA\, 
FWHM resolution from 3838--5423\AA. The red channel observations 
were done with the R1200 l/mm grating in first order, yielding 1.3\AA\, 
FWHM resolution over 6205--7157\AA. The five nights of bright 
time were predominantly clear to partly cloudy. Signal-to-noise ratios of 
$\sim$50-200 per resolution element were typically reached with integration times 
of 120-720 s. Flat-fields and bias-frames were observed at the beginning and 
end of each night. NeAr $\lambda$-calibration arcs and spectrophotometric 
standards were 
observed every few hours. The spectra were reduced using standard IRAF 
routines. In order to remove low-order chromatic effects from the 
band-ratio measurements, we spectrophotometrically calibrated all of the 
target spectra using 2 standard stars from \citet{Hamuy94}.
A total of 118 program stars (\S\ref{z99} and \S\ref{rasstycho})
and 20 MK spectral standards (\S\ref{sptstandard}) were observed.
The major stellar absorption features of one of the single standard 
G stars were shifted to a zero-velocity wavelength scale. 
The spectra of all of the stars were then cross-correlated
against this standard star using the IRAF task {\it fxcor}, and then 
shifted to the common, rest-frame wavelength scale. This was done
to ensure proper identification of weak lines, as well as to 
make sure that the band-ratio measurements were sampling the same
spectral range in each stellar spectrum.

\section{Analysis of Spectra}

\subsection{Spectral Types and Luminosity Classification\label{spt}}

\subsubsection{Standard Stars \label{sptstandard}}

We observed 20 spectral standards including dwarfs
and subgiants (luminosity classes IV and V) and a few giants (III).
A summary of their properties is listed in Table \ref{tab_stan}.
To permit quantitative examination of trends in the strengths of 
spectral features, as well as interpolation between spectral types, we 
adopt the numerical subtype scaling of Keenan (1984; i.e. here listed 
as ``SpT'', where G0 = 30, G2 = 31, K0 = 34, etc.). All of the standard 
stars are classified on the MK system by \citet{Keenan89}, except for HR 
7061 \citep{Garcia89}. Table \ref{tab_stan} also lists their spectral types 
as given in the Michigan Spectral Survey atlases of N. Houk (e.g. 1978). The 
sample standard deviation of a linear fit between the Keenan and Houk spectral 
types for dwarfs and subgiants, on Keenan's subtype scale, is $\sigma$(SpT) = 
0.6 subtypes. The $\sim$0.6 subtype uncertainty probably represents the best 
that can be done using visual spectral types determined by different authors.

The adopted spectral types are those of Keenan \& McNeil's, however the 
luminosity classification was verified (and some times changed) by virtue of 
(1) position of the stars on a color-magnitude diagram based on 
{\it Hipparcos} data, (2) position in a temperature vs. \ion{Sr}{2} 
$\lambda$4077/\ion{Fe}{1} $\lambda$4071 ratio diagram (see \S \ref{quantspt}, 
Fig. \ref{fig:srfe}), and (3) published \logg\, estimates. Although changing 
the classification of some standards may appear 
imprudent, the H-R diagram positions, Sr/Fe line ratio, and derived \logg\, values 
\citep{Cayrel01} {\it are all consistent with our new adopted luminosity 
classes}\footnote{{\it Hipparcos} data also led \citet{Keenan99} to revise the 
luminosity classes of a few giant star standards at the half luminosity class
level.}. In every case the difference was only half of a luminosity 
class, and only 5/20 of the stars were changed. Notes on the revised 
luminosity classifications are given in Appendix B. 

\placetable{tab_stan}

\subsubsection{Visual Classification\label{visspt}}

The blue spectrum of each star was assigned a spectral type visually by E.M. 
through comparison with the standards in Table \ref{tab_stan}. In order 
to distinguish subtypes, we focused on several features such as the G band 
($\lambda$4310), \ion{Ca}{1} $\lambda$4227, \ion{Cr}{1} $\lambda$4254 and 
nearby Fe lines, and the Mg b lines $\lambda$5167, $\lambda$5173, 
$\lambda$5184. Balmer lines were ignored due to possible chromospheric emission. 
After making an initial guess through comparison with a 
wide range of spectral types, a final visual spectral type was assigned 
through comparison to standards within $\pm$2 subtypes of the initial guess. 

To test the accuracy of our visual classification, we compared our spectral 
types to those of quality 1 or 2 in the Michigan Spectral Survey. The average 
difference is not significant: --0.4\,$\pm$\,0.6\,(1$\sigma$ sample standard
deviation) subtypes (on 
Keenan's scale). The 12 {\it Hipparcos} stars were later visually typed a 
second time. Between the two estimates for each star, we estimate that the 
1$\sigma$ uncertainty in our visual spectral types is 0.6 subtypes. 
This is comparable to the dispersion between the Keenan and Houk spectral 
types for the standards themselves.

\subsubsection{Quantitative Spectral Type Estimation \label{quantspt}}

A two-dimensional quantitative spectral type (subtype plus luminosity class) 
can be estimated using integrated fluxes over narrow bands sensitive to 
temperature and surface gravity. We tested various ratios defined by 
\citet{Malyuto97} and \citet{Rose84} for this purpose, as well as 
from Gray's spectral atlas (2000). 
In testing band-ratios as temperature indicators for our 
standards, we noticed that some had slight surface gravity dependencies. 
A surface gravity dependence in our temperature indicators could 
systematically affect our \teff\, estimates. We first discuss our
surface gravity indicator and then define our temperature estimators 
using only subgiant and dwarf standards, thus mitigating the effects of 
surface gravity.

The most widely used surface gravity diagnostic for G and K stars is the ratio 
between \ion{Sr}{2} $\lambda$4077 and nearby Fe lines 
\citep[e.g.][]{Keenan76,Gray00}. In thin and thick disk dwarfs, the 
abundance ratio [Sr/Fe] is within $\sim$0.1 dex of solar for most stars 
\citep{Mashonkina01}. A quantitative surface gravity (luminosity class) indicator 
was established by \citet{Rose84} from low-resolution spectra using the maximum 
absorption line depth for \ion{Sr}{2} $\lambda$4077 and the average for 
the atomic Fe $\lambda$4045 and $\lambda$4063 lines. We measure the fluxes 
in 3\,\AA\, bands centered on the \ion{Sr}{2} $\lambda$4077 line and the 
\ion{Fe}{1} $\lambda$4071 line. Ratios between the $\lambda$4077 line and 
the other nearby Fe lines ($\lambda$4045, $\lambda$4063) did not distinguish 
subgiants and giants.

For a temperature estimator, we adopted Index 6 of \citet{Malyuto97} 
($\lambda\lambda$5125-5245/$\lambda\lambda$5245-5290) hereafter referred to 
as ``MI6''. The temperature sensitivity of this indicator largely reflects 
differing amounts of line-blanketing in these two wavelength regimes --
mainly by the Mg b lines ($\lambda$5167, $\lambda$5173, and $\lambda$5184), 
and many Fe lines (e.g. \ion{Fe}{1} $\lambda$5270). Although the Mg b lines 
are somewhat surface gravity sensitive, within the \logg\, and \teff\, regime of 
our standards and program stars, the temperature sensitivity is dominant.
The difference in central wavelength between the two bandpasses is
only 82\AA, and the effects of reddening are negligible \citep{Mathis90}. 
For a temperature indicator, we fit a low-order polynomial to 
MI6 vs. SpT for the dwarf and subgiant standard stars (see Appendix C) which 
has a 1$\sigma$ sample standard deviation of 0.6 subtypes.

Fig. \ref{fig:srfe} plots the temperature-sensitive MI6 index versus our 
surface gravity discriminant (\ion{Sr}{2} $\lambda$4077/\ion{Fe}{1} $\lambda$4071).
The dwarf standards form a very narrow sequence in Fig. \ref{fig:srfe}, 
{\it confirming the lack of cosmic scatter in [Sr/Fe] values among field 
stars and the insensitivity of \logg\, to spectral type for G-K dwarfs}. 
The polynomial fit to the dwarf data is given in Appendix 
\ref{appendix_poly}. There is a gap between the dwarf and subgiant loci 
between $\sim$1-2.5$\sigma$ (sample standard deviation) of the dwarf locus polynomial, and we 
set the subgiant/dwarf separation at 2$\sigma$. We classify stars within 2$\sigma$ 
of the solid dwarf line in Fig. \ref{fig:srfe} as dwarfs (4 of 96
RASS-ACT/TRC stars, 4 of 20 HIP stars), and three stars near the giant 
locus (TYC 8992-605-1, HIP 68726, and HIP 74501) as giants. We classify the 
rest as subgiants.


\citet{Gray00} suggests \ion{Y}{2}~$\lambda$4376/\ion{Fe}{1}~$\lambda$4383 as a 
surface gravity indicator for late-G stars using low-resolution spectra. 
From the solar spectral atlas of \citet{Wallace98}, it 
appears that Gray's low-resolution \ion{Y}{2}\,$\lambda$4376 feature is actually a blend of 
several lines of nearly equal strength. In order to test the properties of 
this band-ratio, we measure the flux in 3\AA\, windows centered on 
wavelengths 4383.6\AA\, and 4374.5\AA. Plotting this ratio against spectral 
type for the standard stars showed a very tight locus for the dwarfs, 
however luminosity classes IV-V, IV, and III were indistinguishable 
from the dwarfs and each other. We found this ratio unsuitable for the purposes of 
luminosity classification of our targets, but we find it to be an excellent 
temperature estimator for FGK dwarfs, subgiants, and giants. Among the
20 standards, the measurement of the $\lambda$4374/$\lambda$4383 band-ratio vs. 
spectral type gives a tight correlation (sample standard deviation 
1$\sigma$ = 0.6 subtypes). 
We adopt the $\lambda$4374/$\lambda$4383 band ratio as our third, independent 
estimator of spectral type (polynomial fit is given in Appendix \ref{appendix_poly}).

\subsubsection{Final Spectral Types}

The three temperature-type estimates agree well for the majority of
the program stars. The mean difference between the MI6 and visual spectral types 
is 0.7 subtypes. The mean difference between the $\lambda$4374/$\lambda$4383 
band-ratio types and the visual types is 0.6 subtypes. We 
calculate a mean spectral type and standard error of the mean using the three 
classifications. The mean is unweighted since all three relations appeared 
to have 1$\sigma$ sample standard deviations of $\approx$0.6 subtypes in their 
accuracy. The average standard error of the mean is 0.5 subtypes. 
We believe that using multiple 
techniques mitigates the effects that rapid rotation, binarity, 
etc. can introduce into visual classification alone. The spectral types are 
listed in Tables \ref{tab_rasstyc_pms}, \ref{tab_hip_pms} and \ref{tab_others}.

\subsection{Additional Spectroscopic Diagnostics}

\subsubsection{Chromospheric H$\alpha$ Emission \label{Halpha}}
 
Medium-to-low resolution spectra of chromospherically active stars show the 
H$\alpha$ line to be partially filled-in, or even fully in emission. 
We measure the EW of the entire H$\alpha$ feature; our resolution is 
insufficient to separate the ``core'' chromospheric emission from the 
photospheric H$\alpha$ absorption line. A significant number of our stars 
show H$\alpha$ emission (19\% of the \rasstyc\, G-K type stars). We 
characterize our targets stars as chromospherically active or inactive through
comparing the H$\alpha$ EW to that of standards of identical
spectral type. Fig. \ref{fig:ha} shows the EW(H$\alpha$) data
for our targets and standard stars. Stars more than 2$\sigma$ 
above the dwarf/subgiant EW(H$\alpha$) relation (a quadratic regression; see 
Appendix \ref{appendix_poly}) are considered to be active. The stars with 
H$\alpha$ in emission (negative EWs) have an ``e'' appended to their 
spectral types in Tables \ref{tab_rasstyc_pms}, \ref{tab_hip_pms} and 
\ref{tab_others}. The H$\alpha$ EWs for each star are also listed in these 
tables.  


\subsubsection{\ion{Li}{1} $\lambda$6707 Equivalent Width \label{lithium}}

The presense of strong Li absorption in the spectra of late-type stars is a 
well-known diagnostic of stellar youth. Because of the extended timescale for
significant Li depletion in stars of $\sim$1 M$_{\odot}$, {\it strong Li 
absorption is necessary, but not sufficient indicator of pre-MS nature 
for G stars}. However it is a powerful age descriminent when combined with 
our surface gravity indicator.

Many studies have shown that the equivalent width (EW) of \ion{Li}{1} 
$\lambda$6707 can be overestimated at low spectral resolution \citep[e.g.][]{Covino97},
especially for G-K stars. With a resolution of 1.3\AA, we consider our EWs to be 
approximate. The EWs were measured with Voigt profiles in the IRAF routine {\it splot}. 
The continuum level was estimated from nearby pseudo-continuum peaks. 
We subtract the contribution from the neighboring \ion{Fe}{1} 
$\lambda$6707.4\AA\, feature using the prescription of \citet{Soderblom93}.
In order to test the validity of our Li EWs, we divided several of our Li-rich
targets by standard stars of the same spectral type. The ratioed spectra 
exhibit only a major absorption feature at $\lambda$6707. The division
also removes the effects of blending by Fe lines (assuming the same stars 
have similar EWs). The EWs of this feature in the divided spectra corresponded 
well with our previous measurements, however the uncertainties in the EW appear
to be $\sim$20-50 m\AA\, (with the maximum value being for spectroscopic binaries).

Fig. \ref{fig:li} shows the \ion{Li}{1} $\lambda$6707 EWs 
for our RASS-ACT/TRC and {\it Hipparcos} targets, separated according to their luminosity
class (\S\ref{quantspt}). Effective temperatures (\teff) come from 
the final spectral type (see \S\ref{stelpar}). Most points lie above the 
\ion{Li}{1} $\lambda$6707 EWs that characterize young open clusters, plotted as 
low-order polynomial fits for the
IC 2602 \citep[30 Myr;][]{Randich97}, Pleiades \citep[70-125 Myr;][]{Soderblom93,Basri96}, 
and M34 clusters \citep[250 Myr;][]{Jones97}.
The comparison is not completely fair, however, since the cluster ZAMS
stars will be roughly 10\% less massive than the corresponding pre-MS stars.
Even if most of our program stars were older ZAMS stars, they still would
be Li-rich compared to stars in the well-studied open clusters. 
We select as ``Li-rich'' those stars above the solid line in Fig. \ref{fig:li}.
Considering the uncertainties in our EW(Li) measurements, and the lack of any other
$\sim$10-20 Myr-old pre-MS G-K-type stellar samples with which to compare, 
we are not compelled to subdivide our sample further.
We will reserve a more detailed investigation of the Li abundances 
for a future high-resolution spectral study. For the present, we are content
to have demonstrated that we have identified a population which appears
to be more Li-rich than ZAMS stars.   


\section{Defining the PTTS Sample \label{nature}}

\subsection{Membership Status \label{membership}}

Our survey was designed to identify the pre-MS G and K-type stars in the Sco-Cen
OB association. We classified the late-type stars according to their positions
in Figs. \ref{fig:srfe}, \ref{fig:ha} and \ref{fig:li} 
(Table \ref{tab_class}). We consider the 110 stars (85/96 RASS-ACT/TRC and 16/30 
{\it Hipparcos}) classified as ``Li-rich'', 
``subgiant'', and ``active'' as bona fide post-T Tauri stars (``pre-MS''). 
Li-rich stars with subgiant surface gravities and H$\alpha$ EWs similar to 
the standard field stars (i.e. ``inactive'') are called ``pre-MS?''. 
Only 3 of the RASS-ACT/TRC stars, and 6 of the HIP stars
are classified as ``pre-MS?''. 
The lone object with giant-like surface gravity in the RASS-ACT/TRC X-ray-selected 
sample (TYC 8992-605-1) is Li-rich, and we also classify it as a pre-MS PTTS.
The 9 ``pre-MS?'' stars were included in our statistics concerning
the star-formation history and disk-frequency of the sample
(\S\ref{results}) for a total of 110 candidate lower-mass members of
the LCC and UCL subgroups. All 13 stars selected in the \rasstyc\, sample
which overlaped with dZ99's membership lists were found to be pre-MS candidates.
Our \rasstyc\, sample (including the 13 dZ99 stars also selected) 
yielded a pre-MS hit-rate of (88+13)/(96+13) = 93\%. 
Of the 30 dZ99 candidates we observed, 22/30 (73\%) are classified as pre-MS or
``pre-MS?''. The numbers of stars by membership class are listed in 
Table \ref{tab_class}. Pre-MS stars in the {\it Hipparcos} 
sample are listed in Table \ref{tab_hip_pms}, and those in the RASS-ACT/TRC
sample are given in Table \ref{tab_rasstyc_pms}.
Li-rich stars with dwarf-like surface gravity (N = 5) 
were considered young main sequence field stars (``ZAMS''), 
and are listed along with other interlopers in Table \ref{tab_others}.

\placetable{tab_class}

\placetable{tab_hip_pms}

\placetable{tab_rasstyc_pms}

\placetable{tab_others}

\subsection{Sample Contamination}

The primary contaminants one would expect from an X-ray- and 
proper motion-selected sample are X-ray-luminous ZAMS stars (ages $\approx$ 
0.1-1 Gyr). Field ZAMS stars could occupy the same region of UVW 
velocity space, and be selected in our study by virtue of their proper motions
and X-ray emission. However, our selection of candidate Sco-Cen members
utilizes a surface gravity criterion which should minimize contamination.
Even if our surface gravity indicator was in error, we claim ZAMS stars do not
dominate our sample. Field ZAMS stars exhibit a large spread in Li EWs (especially 
for the late-G and early-K stars), however 
this is not observed in Fig. \ref{fig:li}.
The star just below the ``Li-rich'' line in Fig. \ref{fig:li} (TYC 7318-593-1; 
G9, EW(Li) $\simeq$ 150 m\AA) happens to be the sole \rasstyc\, star with 
inferred \logg\, {\it higher} than that of the dwarf standards in 
Fig. \ref{fig:srfe}. We consider TYC 7318-593-1 to be a field
ZAMS star candidate due to its intermediate Li strength and high surface gravity. 

We can rule out most of the pre-MS candidates being
Li-rich {\it post-MS} stars. Based on the surveys of Li abundances in
field subgiants by \citet{Randich99} and \citet{Pallavicini87}, we do not expect to 
find any post-MS subgiants with EW($\lambda$6707) $>$ 100 m\AA. 
Even if our measured EWs for the \ion{Li}{1} $\lambda$6707 line are over-estimated due
to low spectral resolution, the overestimate would have to be greater than
a factor of two to reconcile our sources with even the most Li-rich subgiants
found in the Randich et al. survey. Our spectral analysis suggests that the majority 
of our sample stars are both Li-rich and above the main sequence (i.e. pre-MS).

Could some of our stars be {\it post}-MS chromospherically active binaries
(CABs) or RS CVn systems? The light from an RS CVn system would be dominated
by a rapidly rotating, evolved (subgiant) primary. 
Only six of our targets are Li-poor subgiants
(HIP 63797, 81775, TYC 8293-92-1, 7833-1106-1, 7858-526-1, and 8285-847-1). 
The first three appear to be normal subgiants. 
TYC 7833-1106-1 is possibly a spectroscopic binary. TYC 7858-526-1 has
a wide, broad H$\alpha$ absorption line. It appears to be a multiple 
late-F star (we classify it as F8.5; \citet{Houk82} classify it as F5), so the star 
could hide a cosmic Li abundance due to the 
increased ionization of \ion{Li}{1} in F stars (and correspondingly lower EW(Li)). 
The system could be a legitimate member, but we exclude 
it from the pre-MS sample. The subgiant TYC 8285-847-1 (HIP 69781 = V636 Cen)
is probably a CAB. It is a previously known
grazing, eclipsing binary \citep[e.g.][]{Popper66} and its saturated X-ray emission 
argues for being a true CAB. Finally, the Li-rich star TYC 8992-605-1
(star \#19; K0+III) is the only \rasstyc\, star that appears in the giant regime of 
Fig. \ref{fig:srfe}. The star is an obvious spectroscopic binary of nearly equal mass.
We believe this star is probably a pre-MS binary, and include it in our ``pre-MS?'' sample.
It appears that CABs are a negligible contaminant
when using X-ray and kinematic selection in tandem with medium dispersion
spectroscopy to identify pre-MS populations.

\subsection{Sample Completeness \label{imf}}

We can make a rough estimate how many stars our selection 
procedure should have detected by counting the number of massive 
Sco-Cen members in a certain mass range, and assuming an initial mass function. 
We assume a complete membership census within a limited mass range (the revised
B-star {\it Hipparcos} membership from dZ99), 
and then extrapolate how many stars we should have seen in our survey. 
We produce a theoretical H-R diagram for the subgroups' B stars (discussed 
at length in \S\ref{turnoffage}), and calculate masses from the evolutionary tracks of 
\citet{Bertelli94} (Z\,=\,0.02). We choose 2.5\,\msol\, as our lower mass boundary
(roughly the lower limit for B stars), and adopt 13\,\msol\, as the upper mass boundary
(slightly higher than the highest inferred mass from the main sequence members). 
In this mass range, we count 32 LCC members and 56 UCL members.
We use a \cite{Kroupa01} IMF to predict how many low-mass stars might belong to
the OB subgroups. Down to the hydrogen-burning limit (0.08\,\msol), a total population 
of 1200$^{+200}_{-300}$ stars in LCC and 2200\,$\pm$\,300 stars in UCL is
predicted (Poisson errors)\footnote{For low-number statistical uncertainties, we 
use the 1$\sigma$ values 
from \citet{Gehrels86} throughout.}. Between 1.1-1.4\,\msol, the mass range of a 15 
Myr-old population that
our survey can probe (see \S\ref{premsage} and Fig. \ref{fig:hrd}), the Kroupa IMF 
predicts a population of 29$^{+6}_{-5}$
stars in LCC, and 51$^{+8}_{-7}$ stars in UCL. 
In this mass range, our survey detects 36 pre-MS stars in LCC
and 40 pre-MS stars in UCL. The number of observed pre-MS stars with
1.1-1.4\,\msol\, corresponds to +1.1$\sigma$ and 
-1.6$\sigma$ of the predicted number, for LCC and UCL respectively. 
This suggests that our survey is fairly complete for LCC, but 
we might be missing $\sim$10 members with masses of 1.1-1.4\,\msol\, 
in the more distant UCL subgroup if the
subgroup mass function is consistent with the field star IMF. The
missing members of the UCL OB subgroup could be X-ray
faint (L$_X$ $\leq$ 10$^{30.2}$ \ergs) stars which we were capable
of detecting in the closer LCC subgroup. The IMF
extrapolation does suggest that we have likely found at least 
the majority of stars in this mass range in both OB subgroups
(if not a complete census for LCC) and that our samples
are representative of the total population. 

\section{The H-R Diagram \label{hrd}}

In order to investigate the star-formation history of the LCC and UCL 
OB subgroups, we convert our observational data (spectral
types, photometry, distances) into estimates of temperature and luminosity. 
We then use theoretical evolutionary tracks to infer ages and masses
for our stars.

\subsection{Photometry \label{phot}}

The primary sources of photometry for our sample of association member
candidates are the Tycho-2 catalog \citep[][]{Hog00TYC2a,Hog00TYC2b}
and 2MASS working database. However, the Tycho and 2MASS bandpasses are 
non-standard, and must be converted to standard photometric systems to enable 
comparison with intrinsic colors of normal stars and the interstellar 
reddening vector. To convert the Tycho photometry to the Johnson
system, we fit low-order polynomials to the data in Table 2 of 
\citet{Bessell00} (relations given in Appendix\,\ref{appendix_poly}). 
A caveat is that Bessell's calibrations are for B--G dwarfs and K--M giants. 
The majority of our stars appear to be pre-MS G--K stars, whose intrinsic colors 
should more closely match those of dwarfs rather than giants. To convert the 
2MASS $JHK_s$ data to the system of \citet{Bessell88}, we use the conversions 
of \citet{Carpenter01}. The original optical and near-IR photometry for our 
target stars is given in Tables \ref{tab_rasstyc} and \ref{tab_hip}.

\subsection{Temperature Scale \label{stelpar}}

To fix stellar properties as a function of spectral type,
we adopt relations (i.e. intrinsic colors, BCs) from Table A5 of \citet{Kenyon95}.
Previous studies have shown that colors and BCs as a function of \teff\, are
largely independent of surface gravity over the range of interest
for this study \citep[e.g.][]{Bessell98}. However, \teff\, decreases with lower 
\logg\, for FGK stars. After some investigation (see Appendix \ref{appendix_teff}), 
we decided to adopt the dwarf \teff\, scale of \citet{Schmidt-Kaler82} 
(which \citet{Kenyon95} also use) with a -35\,K offset to account for the effects of 
lower \logg\, in our sample stars. 
The scatter in published dwarf \teff\, scales is 60\,K (1$\sigma$) among 
G stars, so while the shift is systematic, its magnitude is of the order of
the uncertainties. The uncertainties in \teff\, given in column 9 of 
Tables 5 and 6 include the uncertainty in spectral type and the scatter in
published \teff\, scales. The typical 1$\sigma$ uncertainties in \teff\,
for the pre-MS stars is $\approx$100\,K.

\subsection{Secular Parallaxes \label{dist}}

All of the stars in our sample have published proper motions,
but only a few dozen have trigonometric parallaxes measured by 
{\it Hipparcos}. The stars are distributed over
hundreds of square degrees of sky, and inhabit stellar associations
which are tens of parsecs in depth. Adopting a standard distance
for all of the stars in the association introduces unwanted scatter
in the H-R diagram. With accurate proper motions available, we calculate individual 
distances to the pre-MS candidates using moving cluster or ``secular'' 
parallaxes \citep[e.g.][]{Smart68}. We adopt the equations
and formalism of \citet{deBruijne99scocen}, as well as his space motions and convergent
points for the LCC and UCL OB subgroups.
The uncertainties in the secular parallaxes are dominated by the uncertainties
in the proper motion ($\sigma_\pi$ $\propto$ $\sigma_\mu$), but contain
a term added in quadrature accounting for a projected 1\,km s$^{-1}$ internal 
velocity dispersion \citep[see \S4 of ][]{deBruijne99scocen}. 
The secular parallax is only meaningful if the
star is indeed a member of the group. Our spectroscopic survey
has confirmed that most of the candidate stars are legitimately pre-MS, 
and that they are most likely members of the OB subgroups. 
Secular parallaxes for older, interloper stars are meaningless 
and ignored. In Tables \ref{tab_hip_pms} and \ref{tab_rasstyc_pms}, 
we list the secular parallaxes and membership probabilities for the pre-MS stars in our 
survey. We calculate membership probabilities P$_1$ and P$_3$ (using
formulae 4 and 6 from dZ99), which have assumed internal velocity dispersions of 
1\,\kms\, \citep{deBruijne99scocen} and 3\,\kms (dZ99), respectively.

The robustness of our method can be illustrated (Fig. \ref{fig:dist})
by comparing the secular parallaxes ($\pi_{sec}$) to the {\it Hipparcos} 
trigonometric parallaxes ($\pi_{HIP}$). The uncertainties are typically 1-2 
mas for the {\it Hipparcos} parallaxes, and 0.5-1 mas for our secular parallax 
estimates. The secular and trigonometric parallaxes agree quite 
well for the few pre-MS stars in our sample for which {\it Hipparcos} measured
the parallax. The secular parallaxes yield distance uncertainties of 
$\sim$5-15\,\% for most of the pre-MS stars. 


\subsection{Luminosities \label{calclum}}

With five-band photometry, a temperature / spectral type estimate, and 
a secular parallax, we calculate stellar luminosities for the pre-MS candidates. 
We adopt the absolute bolometric magnitude of the Sun (\mbol$_{\odot}$ = 4.64)
from \citet{Schmidt-Kaler82}. In order to compromise between the uncertainties 
in luminosity due to reddening, photometric uncertainties, and possible K-band excess, 
we calculate the \mbol\, using the dereddened 2MASS J magnitude. 
We estimate the visual extinction from a weighted mean of
A$_V$ estimates from the color excess in $\bvj$ and $\vj$. We took the E$(\bvj)$
formula from \citet{Drilling00} and the value of A$_J$/A$_V$ (= 0.294) was taken 
from the near-IR extinction law of \citet{Mathis90} for a central wavelength 
1.22\,$\mu$m. The reddening A$_J$ typically ranged from 0 to 0.35 mag with formal 
uncertainties of $\sim$0.1 mag. The typical uncertainty in \logl\, for the
pre-MS candidates is $\approx$0.08 dex. With the luminosities and
X-ray fluxes from the RASS BSC catalog \citep{Voges99}, we calculate
the ratio of X-ray/bolometric radiation for the stars with X-ray counterparts.
The derived values of \lxlbol\, are in the range of $10^{-2.8}-10^{-3.8}$, 
indicating coronal X-ray emission elevated above 
most ZAMS G-type stars \citep[e.g. Pleiads;][]{Stauffer94}. 

\subsection{Evolutionary Tracks}

In order to infer theoretical masses and ages from our pre-MS candidates, we
use the evolutionary tracks from \citet{D'Antona97} (DM97; Z = 0.02, x$_D$ = 
2$\times$10$^{-5}$), \citet{Siess00} (SDF00; Z = 0.02), and \citet{Palla01} (PS01). 
Ages and masses for a given \logt\, and \logl\, were calculated using an 
interpolation algorithm. Given the mean observational errors 
($\sigma$(log\teff,\logl) = 0.007, 0.078 dex for LCC Pre-MS stars, and 
$\sigma$(log\teff,\logl) = 0.009, 0.084 dex among UCL Pre-MS stars), we estimate the 
isochronal age uncertainties for an individual star to be approximately 4, 5, 7 Myr 
(DM97, PS01, SDF00) in LCC, and 4, 5, 5 Myr (DM97, PS01, SDF00) in UCL, as illustrated 
in Fig. \ref{fig:hist2}. The uncertainties in the interpolated masses are 0.1 
M$_{\odot}$ for all three sets of tracks. Fig. \ref{fig:hrd} shows the H-R diagram for 
the pre-MS candidates overlayed with the evolutionary tracks of DM97.


\section{Results \label{results}}

The ages of the low-mass population of LCC and UCL have not been 
estimated before, though \cite{deGeus89} and \cite{deZeeuw85} give
main sequence turn-off ages. 
In \S\ref{premsage} we estimate the pre-MS ages for the two subgroups,
and put an upper limit on the intrinsic age spread. In \S\ref{turnoffage} we
calculate new turn-off ages for the subgroups using early B stars from 
the revised {\it Hipparcos} membership lists of dZ99.

\subsection{Pre-MS Ages and Age Spread \label{premsage}}

The H-R diagram for our ``pre-MS'' and ``pre-MS?'' stars is shown in 
Fig. \ref{fig:hrd}, overlayed with the evolutionary tracks of \citet{D'Antona97}. 
The temperatures and luminosities of the pre-MS stars are given in 
Columns 9 and 10 of Tables \ref{tab_rasstyc_pms} and \ref{tab_hip_pms}, along
with their inferred masses and ages (columns 15 through 17).
One notices immediately that the bulk of isochronal
ages are in the range of $\sim$10-20 Myr. The age range is nearly identical
for both groups. To assess the effects of our magnitude limit in biasing our
mean age estimates, in Fig. \ref{fig:agestat} we plot the 
mean pre-MS age (with standard errors of the mean) for the pre-MS 
subgroup samples as a function of 
minimum \logt\, cut-off. The magnitude bias of our survey is clearly apparent:
the mean age systematically decreases when stars with \logt\, $<$ 3.73 are included
in the calculation. In calculating the pre-MS ages of the OB subgroups,
we explicitly omit the pre-MS stars with \logt\, $<$ 3.73 (30\% of our sample). 
This temperature threshold intersects our magnitude limits at 
ages of $\sim$25 Myr for stars of 1\,\msol\, on the DM97 tracks. 
That the lines in Fig. \ref{fig:agestat} are nearly flat for \logt\, $>$ 3.73, 
suggests that (detectable) stars with ages of $>$25 Myr are not a significant 
component of either subgroup (also see discussion in \S\ref{before}). 


Fig. \ref{fig:hist} displays histograms of the isochronal ages
for the pre-MS stars in the LCC and UCL subgroups derived using DM97 and SDF00
evolutionary tracks. These tracks represent the extrema in age estimates for our
sample (DM97 is youngest, SDF00 is oldest). The 1$\sigma$
age dispersion among the unbiased samples (\logt\,$>$\,3.73) is 5-9 Myr for both
groups. If we remove the known spectroscopic binaries (see notes in Tables 
\ref{tab_rasstyc_pms} and \ref{tab_hip_pms}), the age dispersions are
4-8 Myr. Because there may be additional unresolved binaries, this {\it observed} 
age spread places an upper limit on the {\it intrinsic} age spread. 
As illustrated in Fig. \ref{fig:hist2}, the individual H-R diagram positions of
the pre-MS samples have \logt\, and \logl\, errors which fold onto the 
evolutionary tracks with age uncertainties of 4-7\,Myr. From this analysis,
we conclude that the {\it intrinsic} 1$\sigma$ age dispersions in each subgroup 
must be less than 2-8\,Myr (i.e. $\sim$2/3rds of the star-formation took place in 
$<$4-16\,Myr). Using the DM97 pre-MS ages, which agree best with the turn-off
age estimates (\S\ref{turnoffage}), we find intrinsic 1$\sigma$ age dispersions of 
2 Myr (LCC) and 3 Myr (UCL).{\it This implies that 68\% of the low-mass 
star-formation took place within $<$4-6 Myr, and 95\% within $<$8-12 Myr in the 
OB subgroups}. 
Our observational uncertainties and lack of knowledge about 
the unseen binarity of the pre-MS sample stars do not allow us to constrain the 
age spread more precisely than this. The mean age estimates for our 
unbiased pre-MS samples (log\teff $>$ 3.73, SBs removed) are shown in Table 
\ref{tab_age}. Counter to previous studies, we find that LCC is slightly older than UCL 
by 1-2\,Myr (at 1-3$\sigma$ significance), independent of which evolutionary 
tracks we use. From Fig. \ref{fig:hist}, we also conclude that star-formation
ceased approximately $\sim$5-10 Myr ago in the subgroups. 

\placetable{tab_age}

\subsection{Turn-off Ages\label{turnoffage}}

De Geus et al. (1989) published the most recent age estimates for the LCC and UCL groups, 
but in light of the new {\it Hipparcos} distances
and subgroup membership lists, we feel it is worthwhile reevaluating the subgroups' turn-off
ages. We construct a theoretical H-R diagram for the B-type subgroup members of UCL and
LCC listed both in Table C1 of dZ99 and Tables
A2 and A3 of \citet{deBruijne99scocen}. Several of the ``classical''
\footnote{``Classical'' members are early-type stars which were included in 
Sco-Cen membership lists before the {\it Hipparcos} studies of dZ99.} 
members rejected as members by {\it Hipparcos} from dZ99 are included as well. 
For input data, we use the following
databases in order of availability: (1) $ubvy\beta$ photometry 
from the database of \cite{Hauck97}, (2) $UBV$ photometry from \citet{Slawson92}, 
and (3) $UBV$ photometry from SIMBAD. For distances we use the secular 
parallaxes ($\pi_{sec}$) given in column 4 of Tables A2 and A3
of \citet{deBruijne99scocen} when available, or the {\it Hipparcos} parallaxes ($\pi_{HIP}$). 
We deredden the stars with $ubvy\beta$ photometry 
to the B-star sequence of \citet{Crawford78} using the prescription
of \citet{Shobbrook83}. For stars with Stromgren photometry, we calculate \teff\, using
the temperature relation of \citet{Napiwotzki93}.
If no $ubvy\beta$ photometry was available, we use
$UBV$ photometry to calculate the reddening-free index $Q$ \citep{Crawford76} to
infer the star's unreddened color.
A polynomial fit to Table 15.7 from \citet{Drilling00} is
used to calculate \teff\, as a function of $(B-V)_{\circ}$. 
The BC versus \teff\, relation of \cite{Balona94} is used for all stars. 
We linearly interpolate between the isochrones from \citet{Bertelli94} 
(Y\,=\,0.28, Z\,=\,0.02, convective overshoot) to infer ages for 
the subgroup B stars. The theoretical H-R diagram is shown in Fig. \ref{fig:postms}. 


UCL has a well-defined MS turn-off composed of the {\it Hipparcos} members
HIP 67464 ($\nu$ Cen; B2IV), HIP 68245 ($\phi$ Cen; B2IV), 
HIP 68282 ($\nu^1$ Cen; B2IV-V), HIP 71860 ($\alpha$ Lup; B1.5III), 
HIP 75141 ($\delta$ Lup; B1.5IV), HIP 78384 ($\eta$ Lup; B2.5IV), and
HIP 82545 ($\mu^2$ Sco; B2IV), as well as ``classical'' members 
(but {\it Hipparcos} non-members) HIP 82514 ($\mu_1$ Sco; B1.5Vp) and 73273 
($\beta$ Lup; B2III)\footnote{ 
Note that the two {\it Hipparcos}
non-members were found to be probable members by \citet{Hoogerwerf00} if the 
long-baseline ACT (HIP 73273) and TRC (HIP 82514) proper motions were used 
instead of the {\it Hipparcos} values.}.
The variable star HIP 67472 ($\mu$ Cen; B2Vnpe) was excluded.
Using the \citet{Bertelli94}
tracks, the mean age of the 7 turn-off {\it Hipparcos} members is
17\,$\pm$\,1\,Myr. Including the two ``classical'' members has negligible
effect on the mean age estimate. Our MS turn-off age estimate for UCL
is slightly older than de Geus et al.'s (14-15 Myr), and is close to the 
mean pre-MS ages that we found in \S\ref{premsage} (15-22 Myr). 

LCC lacks a well-defined turn-off, however we have enough early-B stars
in the middle of their MS phase with which to make an age estimate.
We estimate the age for LCC from the following main sequence B stars:
HIP 59747 ($\delta$ Cru; B2IV), HIP 60823 ($\sigma$ Cen; B2V), 
HIP 61585 ($\alpha$ Mus; B2IV-V), HIP 63003 ($\mu^1$ Cru; B2IV-V), and
HIP 64004 ($\xi^2$ Cen; B1.5V).
The mean age for these five stars is 16\,$\pm$\,1 Myr; similar to what we found 
for UCL, and it agrees well with the younger end of our pre-MS age estimates 
(17-23 Myr). This age estimate is significantly older than previous estimates 
(10-12 Myr), and warrants more critical examination (\S\ref{discussion_lcc}).

The new results yielded by our age analysis of the OB subgroups are:
(1) two-thirds of the low-mass star-formation in each subgroup took place in 
less than a $\sim$5 Myr span (and 95\% took place within $\sim$10 Myr), 
(2) the pre-MS and B-star ages for LCC and UCL are in approximate agreement, 
(3) the B-star subgroup memberships defined by {\it Hipparcos} have ages of 
16\,$\pm$\,1\,Myr and 17\,$\pm$\,1 Myr for LCC and UCL, respectively.
We discuss the implications of these results in \S\ref{discussion}.

\subsection{The Census of Accretion Disks \label{disks}}

An important question both for star and planet formation is the lifetime of accretion 
disks around young stars. Statistics for the frequency of active accretion disks
around low-mass stars come predominantly from near-IR surveys of young associations
and clusters \citep[][]{Hillenbrand99,Haisch01}. The samples of 
low-mass stars surveyed are dominated by embedded associations with ages of $<$3 Myr 
(e.g. Tau-Aur, Cha I, etc.), and older open clusters with ages of 30-100 Myr
(e.g. Pleiades, IC 2602, $\alpha$ Per, etc.). Few well-studied pre-MS stars of
3-30 Myr-old ages have been surveyed. The situation has
recently been slightly ameliorated by the discoveries of the TW Hya association
and $\eta$ Cha cluster \citep[][]{Kastner97,Webb99,Mamajek99}. Yet these samples
are small ($\sim$10-20 stars) and dominated by K and M-type stars
with masses of 0.1-0.8 \msol. Our pre-MS star sample is unique
in its mass ($\sim$1-1.5\,\msol) and age range ($\sim$10-20 Myr), 
so measuring its disk-frequency provides a valuable datum.

Stars with EW(H$\alpha$) $>$10\AA\, in emission are usually
called Classical T Tauri stars (CTTSs), which show spectroscopic signatures
of accretion as well as near-IR excesses \citep[e.g.][]{Hartigan95}. 
Stars lacking the strong H$\alpha$ emission and near-IR excesses are
called Weak-lined T Tauri stars (WTTSs). This can be explained as a correlation
between the presence of magnetospheric accretion columns and an inner accretion 
disk \citep[e.g.][]{Meyer97,Muzerolle98}. Our H$\alpha$ EW measurements
are discussed in \S\ref{Halpha}, and here we quantify the K-band excess of
our targets. We calculate the intrinsic $\jk$ color excess E$(\jk)_{\circ}$ as 
defined by \citet{Meyer97}:

\begin{equation}
E(J-K)_{\circ} = (J-K)_{obs} - (J-K)_{\circ} - A_{V}\times\,(A_{J} - A_{K})
\end{equation}
 
\noindent where $(\jk)_{obs}$ is the observed color and $\jk_{\circ}$ is the intrinsic
color of an unreddened dwarf star of appropriate spectral type \citep{Kenyon95}.
Uncertainties in each quantity were propagated in order to estimate the
signal-to-noise of the intrinsic color excess.  
The distribution of measured E$(\jk)_{\circ}$ values indicate a
systematic offset of a few hundredths of a mag. Our near-IR data set is from the 
2MASS working database, so we suspect that the absolute calibration
or uncertainties in color correction could be responsible and apply a 
small color correction to account for it. After the correction, the 
distribution of E$(\jk)_{\circ}$ values
is symmetric about zero, with a few positive and negative $\sim$2$\sigma$ points. 
There is only one star with a E$(\jk)_{\circ}$ color excess 
with S/N $>$ 2.5: star \#34 = TYC 9246-971-1 
has an intrinsic color excess of E$(\jk)_{\circ}$ = 0.26\,$\pm$\,0.06 implying 
a K-band excess. This star
also happens to be the only CTT identified in our optical spectra
(EW(H$\alpha$) = -39\,\AA). TYC 9246-971-1 (= PDS 66, Hen 3-892) was originally
identified as an emission line star by \citet{Henize76}, and classified
as a CTT in the Pico dos Dias survey of stars in the IRAS PSC catalog
\citep{Gregorio-Hetem92}. By virtue of its position, proper motion,
and spectral characteristics, we find that TYC 9246-971-1 is a
$\approx$8 Myr-old, $\approx$1.2\,\msol\, (DM97 tracks) member of the LCC subgroup. 
Our secular parallax for TYC 9246-971-1 yields a distance of 86$^{+8}_{-7}$\,pc; 
the third nearest of the LCC pre-MS stars in our sample, and among the nearest 
CTTSs known. Only 1/58 (1.7$^{+4.0}_{-1.4}$\%; 1$\sigma$ Poisson)
of pre-MS stars in LCC are classified as bona fide CTTS, along with 
none (0/42) of the pre-MS members of UCL. 
For our accretion disk frequency statistics, we use the isochronal ages
derived from the DM97 evolutionary tracks (which agree well with the turn-off ages), 
and include the entire sample of 110 pre-MS stars (including the cooler stars which bias
the mean to younger ages). 
Only 1/110 (0.9$^{+2.1}_{-0.8}$\%; 1$\sigma$) of 1.3\,$\pm$\,0.2\,(1$\sigma$)\,\msol\, stars with ages 
of 13\,$\pm$\,1\,(s.e.)\,$\pm$\,6 (1$\sigma$) Myr are CTTSs. 
This implies that accretion terminates in solar-type stars within 
the first 15 Myr of their evolution.

\section{Discussion \label{discussion}}

We can address several interesting questions regarding the star-formation history
of Sco-Cen with data from our survey.
Could the Sco-Cen progenitor giant molecular cloud (GMC) have produced
a substantial population of low-mass stars for an extended period ($>$5-10 Myr) 
{\it before} conditions were right 
to form an OB population? Conversely, is there evidence for any low-mass star-formation
{\it after} the bulk of the high-mass OB stars formed?
The OB star-formation in LCC and UCL has apparently destroyed the
progenitor GMC through supernovae and stellar winds 
\citep[e.g.][]{deGeus92,Preibisch00}. However the region is not totally
devoid of molecular gas (e.g. the Lupus complex). 
We will first examine whether there is any evidence
of star-formation prior to the formation of the OB subgroups (\S\ref{before}), 
and then assess the evidence for more recent star-formation in the UCL region (\S\ref{lupus}).
We will address the age of LCC in \S\ref{discussion_lcc}, and discuss
the formation of the subgroups in \S\ref{starformationhistory}
Throughout the discussion, we adopt the DM97 ages, as they agree more closely with 
the turn-off ages than do the SDF00 and PS01 ages.

\subsection{Is There Evidence for Star-Formation Before the Primary
Bursts? \label{before}}

Is our survey sensitive to older stars which may have preceded
the primary star-formation episode? Three pre-MS stars
in our sample have isochronal ages of $>$25 Myr (or undefined as lying
below the ZAMS), however given the uncertainties in \teff\, and \logl\, 
(Fig. \ref{fig:hist2}), even a coeval $\approx$15 Myr-old population would 
be expected to have {\it statistical} outliers. Here we explore three 
possible ways in which older ZAMS stars could have escaped our attention.

One could argue that our surface gravity indicator is biasing our 
sample against identifying ZAMS stars members (if they exist). 
If we disregard surface gravity as a criterion, we gain only 4 more 
\rasstyc\, stars (all between F8.5 and G1), and only {\it one} of those 
would have an isochronal age $>$\,25 Myr (TYC 8222-105-1; $\sim$30 Myr). 
If they were legitimate, older members with real ages of
$>$\,25 Myr, they should also be among the stars with the oldest
{\it isochronal ages} in our sample, which they are not.
This suggests that their secular parallaxes, hence their luminosities,
are unjustified, and that they are not members of the OB subgroups.
Coincidently, TYC 8222-105-1 is one of the earliest type stars in our sample (F8.5),
where our surface gravity indicator has
the least fidelity (Fig. \ref{fig:srfe}). We can state that only 
{\it one} of the Li-rich stars showing dwarf gravity signatures that is 
co-moving with LCC and UCL has an H-R diagram position and gravity suggestive of 
ZAMS status. 

If a significant ZAMS population existed in LCC and UCL, would our magnitude
and X-ray flux limits  allowed their detection?
X-ray surveys of the ZAMS-age clusters IC 2602 and IC 2391 ($\sim$30-50 Myr) by 
\citet{Randich95} and \citet{Patten96} found that late-F and early-G stars ZAMS stars have 
X-ray luminosities of L$_X \simeq 10^{29.0} - 10^{30.5}$ erg s$^{-1}$, with 
L$_X$/L$_{bol}$ ranging from 10$^{-3.0}$ to 10$^{-4.8}$. The X-ray and optical flux limits
imposed by the {\it ROSAT} All-Sky Survey and the Tycho catalog allow us
to detect ZAMS sources with L$_X$/L$_{bol}$ $>$ 10$^{-3.2}$ within
140 pc if they exist. If we adopt the X-ray luminosities of the G stars 
in IC 2602 and IC 2391 as
representative for a $\sim$30 Myr-old population, we should have 
detected roughly one-third of a putative Sco-Cen ZAMS population between
masses of 1 and 1.2\,\msol. We can put a rough
upper limit on the number of $>$25 Myr-old stars in our mass range. 
Assuming that TYC 8222-105-1 {\it is} a ZAMS member, and that its H-R diagram
position is not a statistical fluctuation from the locus of $\sim$15 Myr-old stars,
we detect one ZAMS star with age $>$25 Myr in the mass range 
(1-1.2\,\msol). Accounting for the two-thirds of the ZAMS stars which would have 
undetectable X-ray emission, and extrapolating over a \citet{Kroupa01} IMF,
this implies a population of $\sim$100 stars with masses greater 
than 0.1\,\msol. This is $\leq$10\% of the stellar population predicted 
to exist in each OB subgroup ($\sim$1000-2000; \S\ref{imf}).

Could such ZAMS stars have left the region
we probed? If we postulate that the population was very centrally concentrated
and gravitationally bound until the OB stars destroyed the giant molecular cloud 
(GMC) some $\sim$10 Myr ago \citep{deGeus92}, then a 2 \kms\, motion radially away 
from the subgroup center would have moved the star 20\,pc in the past 10 Myr. 
This distance is the approximate radius of both of the subgroups today (see Fig. 9 of 
dZ99). Hence if an older population was concentrated at center of the 
gravitationally-bound GMC until the high mass stars destroyed the cloud, 
we would find them within the projected boundaries of the subgroups so long 
as they inherited velocities of $<$2 km s$^{-1}$. The kinematic selection procedure 
of \citet{Hoogerwerf00} would have selected such stars, since a large 
velocity dispersion (3 km\,s$^{-1}$) was initially assumed.

We conclude that there is no evidence for significant 
star-formation in the LCC and UCL progenitor giant molecular clouds before 
the primary star-formation episodes. Our findings are consistent with the idea
that molecular clouds form stars over a range of masses, and dissipate within timescales
of $\sim$10 Myr. 

\subsection{On-going Star-Formation? \label{lupus}}

Two obvious sources of young stars may be contaminating the UCL pre-MS sample.
The youngest, unembedded OB subgroup of Sco-Cen is US, 
with a nuclear age of 5-6 Myr \citep{deGeus89}. US borders UCL near
Galactic longitude 343$^{\circ}$, and its space motion and distance are very similar
to that of UCL (dZ99). The Lupus molecular clouds are also
in the western region of UCL (roughly between 
335$^{\circ}$ $<$ $\ell$ $<$ 345$^{\circ}$ and +5$^{\circ}$ $<$ $b$ $<$ 25$^{\circ}$).
The T Tauri star population within the major Lupus clouds was surveyed by 
\cite{Hughes94}, and the region was recently mapped in $^{12}$CO by \cite{Tachihara02}.
Dozens of pre-main sequence stars were identified outside of the main cores
by a pointed {\it ROSAT} survey \citep{Krautter97}, and the All-Sky Survey 
\citep{Wichmann97GB}. The clouds lie at $d$ = 140\,pc \citep{Hughes93},
situated spatially between the US and UCL subgroups of Sco-Cen (both
$d$\,$\simeq$\,145\,pc). 
Fig. \ref{fig:map} illustrates the positions of the primary Lupus clouds, 
the pre-{\it ROSAT} Lupus T Tauri star population, the pre-MS
stars from our survey, and the B-star population of the OB subgroups. 


How does the presence of US and the Lupus molecular clouds (and
their associated T Tauri stars) affect our findings regarding the mean age of 
the UCL subgroup? We split our unbiased (\logt\,$>$\,3.73) ``pre-MS'' and 
``pre-MS?'' members of UCL into two groups using the Galactic longitude line 
335$^{\circ}$ as a division. Most of the molecular cloud mass in the Lupus 
region lies between 335$^{\circ}$ $<$ $\ell$ $<$ 345$^{\circ}$ (see Fig. 2 of 
\citet{Tachihara02}). Using the DM97 tracks, we find that the ``eastern'' UCL pre-MS 
sample surrounding the Lupus
clouds has a mean age of 13\,$\pm$\,1\,Myr, while the ``western'' UCL 
sample is somewhat older (16\,$\pm$\,1\,Myr). The UCL stars
with ages of $<$10 Myr are found in greater numbers near the Lupus
clouds and US border, supporting the idea that our UCL sample is probably contaminated
by more recent star-formation. 
The age estimate of UCL for the stars west of $\ell$ = 335$^{\circ}$ is probably
more representative of the underlying UCL population.

Three of the youngest stars (HIP 81380, TYC 7858-830-1, and TYC 7871-1282-1; 5-9 Myr; 
DM97 tracks) in our entire survey are positioned near a clump of 8 B stars at 
($\ell$, $b$) = (343$^{\circ}$, +4$^{\circ}$). These three pre-MS stars also
have secular parallax distances of $\sim$200\,pc, similar to what 
\citet{deBruijne99scocen} found for the group of B stars. The secular parallaxes may be biased, 
however, if this clump has slightly different kinematics than the average UCL motion. 
This clump may represent substructure within UCL. However \citet{deBruijne99scocen} was unable 
to demonstrate that the clump had distinct kinematics or age. The mean 
{\it Hipparcos} distance of the clump B stars is 175\,pc, with HIP 82514 
($\mu^1$ Sco; B1.5Vp) and HIP 82545 ($\mu^2$ Sco; B2IV) as the most 
massive members. Our identification of 3 new pre-MS stars in the same region with similar
secular parallaxes supports the notion that this may be a separate subgroup.

Some of our pre-MS stars were also identified in the {\it ROSAT} surveys
of the Lupus region by \citet{Krautter97} and \citet{Wichmann97GB} 
(see notes in Table \ref{tab_rasstyc_pms}). The presense of a significant population of 
pre-MS stars outside of star-forming molecular clouds
has been attributed by various authors to be due to one or more of the following: 
(1) slow diffusion (1-2 \kms) from existing molecular clouds \citep[e.g.][]{Wichmann97Lup}, 
(2) ejection from small-N body interactions \citep{Sterzik95}, 
(3) formation {\it in situ} from short-lived cloudlets \citep{Feigelson96}, 
or (4) fossil star-formation associated with the Gould Belt \cite[e.g.][]{Guillout98}. 
Wichmann convincingly showed that most of the young RASS stars in the Lupus region are 
at a distance of around $\sim$150\,pc (similar to previously published distances for the 
Lupus clouds and UCL), and that the stars are roughly 10 Myr-old. Wichmann concludes that
the dispersed pre-MS population is most likely a manifestation of the Gould Belt.
The OB subgroups of Sco-Cen are major sub-structures of the Gould Belt 
\cite[as defined by age and kinematics;][]{Frogel77}, i.e. UCL is the
dominant Gould Belt substructure in the Lupus region. 
We interpret the presence of dozens of pre-MS stars near the Lupus clouds 
to be primarily the low-mass membership of the UCL OB subgroup. 
Our analysis suggests that younger US or Lupus stars are a minor contaminant to our 
UCL sample.

\subsection{Is LCC Older than UCL?\label{discussion_lcc}}

Although our pre-MS and turn-off age estimates for LCC agree rather well, they are 
substantially older (by $\sim$50\%) than previous values. 
The \citet{deGeus89} age estimate (11-12 Myr) appears to hinge primarily on the
H-R diagram position of $\epsilon$ Cen, with $\delta$ Cru, $\alpha$ Mus, and $\xi^2$ 
Cen defining the rest of the turn-off isochrone. The latter three stars were 
confirmed as members by dZ99, however $\epsilon$ Cen (the most massive) 
was rejected. Although our age for $\epsilon$ Cen is consistent with de Geus's,
we omitted it from our LCC age estimate.
If one uses the long-baseline proper motion for $\epsilon$ Cen from the new
Proper Motions of Fundamental Stars (PMFS) catalog \citep{Gontcharov01}, and 
adopt the LCC space motion, convergent point,
and formulae of dZ99 (with $v_{int}$ = 3\,\kms), 
$\epsilon$ Cen has a 100\% membership probability. 
The resulting secular parallax ($\pi_{sec}$\,=\,9.6\,$\pm$\,2.0 mas)
agrees well with the {\it Hipparcos} astrometric parallax ($\pi_{HIP}$\,=\,8.7\,$\pm$\,0.8 mas),
further strengthening the interpretation that $\epsilon$ Cen is a bona fide LCC member.
Including $\epsilon$ Cen with the other 5 turn-off stars discussed in \S\ref{turnoffage}
does not change our turn-off age estimate, however (16\,$\pm$\,1 Myr).
If one ignores the stars with masses less than that of $\epsilon$ Cen, then the 
12-Myr Bertelli isochrone would appear to be an acceptable fit for LCC. Because the turn-off is 
poorly defined, we give equal weight to the next five {\it Hipparcos} members
down the mass spectrum ($\delta$ Cru, $\sigma$ Cen, $\alpha$ Mus, $\mu^1$ Cru, and $\xi^2$ Cen),
which yields an age older than de Geus's.  

$\epsilon$ Cen is one of several ``classical'' LCC early B-type member candidates 
rejected as members of Sco-Cen using the {\it Hipparcos} astrometry. These stars have been 
included in Sco-Cen candidate membership lists on and off over the past half-century: 
HIP 59196 ($\delta$ Cen; B2IVne), HIP 60718A ($\alpha^1$ Cru A; B0.5IV), 
HIP 62434 ($\beta$ Cru; B0.5IV), and HIP 68702 ($\beta$ Cen; B1III). 
These stars are $\sim$10-20\,\msol\, star, with inferred ages of
$\sim$5-15 Myr, and distances of $\sim$100-150\,pc. Such stars are extremely rare, 
and their presense in the LCC region appears to be more than coincidental. 
Are they all LCC members whose {\it Hipparcos} proper motions are 
perturbed due to binarity? {\it All} five systems are flagged (field \#59)
in the {\it Hipparcos} catalog as stars with unusual motions due to either unseen
companions or variability. A kinematic investigation of these 
stars, and their potential membership in LCC is beyond the scope of this study, but
necessary for understanding the global star-formation history of the Sco-Cen region.
Are these stars bona fide members? If so, why are they so much younger than the 
other members (both pre-MS and mid-B stars)? If they are not bona fide members, 
where did they come from? Although our age estimates for the pre-MS 
sample and {\it Hipparcos} early-B members appear to be consistent, the presense of these 
young, B0-B2 classical members ({\it Hipparcos} non-members) hints that the story of 
star-formation in Sco-Cen is more complex than our results reveal. 

\subsection{A Star-Formation History of Sco-Cen?\label{starformationhistory}}

\citet{Preibisch00} reviewed the recent star-formation history of Sco-Cen
($<$5-10 Myr) in the region of US, UCL, and $\rho$ Oph. They present evidence for
external supernovae triggering in the formation of the subgroups. 
They claim that supernovae
shock waves from UCL passed through the US progentior GMC approximately 5 Myr ago,
and caused the cloud to collapse. The US group appears to have had at least
one supernova in the past $\sim$1 Myr, possibly a deceased massive companion to
the runaway O9.5V star $\zeta$ Oph. This supernova contributed to destroying the 
GMC and producing the US superbubble \citep{deGeus92,Hoogerwerf01}. The US subgroup
appears to be currently triggering star-formation in the $\rho$ Oph cloud core.
Here we speculate on the global star-formation history of the Sco-Cen complex. 

The formation of LCC and/or UCL may have been similarly triggered. However it is unclear
if one triggered the formation of the other or vice versa. Our pre-MS age estimates are 
consistent with LCC being slightly older than UCL by a few Myr, though they could be coeval.
What was the origin of these large OB subgroups? The gas associated with the Sco-Cen complex 
appears to be part of the Lindblad Ring, a torus of \ion{H}{1} and molecular clouds hundreds
of pc in radius. It is centered roughly near the $\alpha$ Persei cluster and Cas-Tau 
OB association \citep{Blaauw91,Poppel97}. The young stars that have formed from this gas 
complex (i.e. the Gould Belt) share a systematic expansion consistent with a localized 
origin for the whole complex -- probably an expanding gas shell from a large 
star-formation event \citep[e.g.][]{Moreno99}. The gas associated with Sco-Cen appears 
to be part of a ``spur'' of neutral hydrogen and molecular clouds that runs from near 
LCC (including Coalsack, Musca, and Chamaeleon clouds), through Lupus, Ophiuchus, and 
into the Aquila and Vulpecula Rift regions \citep[see Fig. 3-18 of ][]{Poppel97}. It is 
likely that LCC and UCL were among the first clumps in the Lindblad Ring to collapse 
and form stars \citep[see \S4 of ][]{Blaauw91}, either from self-gravity
or from triggered from external supernovae events. The LCC and UCL regions formed a 
large population of OB stars and their stellar winds and supernovae may indeed have 
triggered the collapse of the US group. The process might continue over the next 
10 Myr as the supernovae from the US and $\rho$ Oph subgroups send shock waves into 
the vast reservoir of atomic and molecular gas associated with the Aquila Rift 
\citep[see \S3.4 of ][and references therein]{Poppel97}. On the other side of 
Sco-Cen, there appears to be little gas westward of LCC until one reaches the Vela 
complex some 400\,pc away. The lack of a sufficient gas 
reservoir probably explains why triggering did not proceed to form OB subgroups west of LCC. 

The Sco-Cen subgroups have formed their own network of superbubbles with radii of 
$\sim$100\,pc \citep{deGeus92}. The superbubbles appear to be largely \ion{H}{1}, 
presumably gas from the progenitor Sco-Cen GMC as well as the swept-up interstellar
medium. In some regions, they are associated with well-known nearby molecular 
cloud complexes: Coalsack, Musca, Chamaeleon, Corona Australis, Lupus, and numerous 
small high Galactic latitude clouds \citep[e.g.][]{Bhatt00}. The Lupus clouds are 
spatially coincident with the western side of the US superbubble, however no 
kinematic analysis has been yet undertaken to determine whether the Lupus clouds 
share in the bubble expansion. The CrA molecular clouds are embedded within 
the UCL superbubble shell, and the space motion of the T Tauri star population is 
moving radially away from UCL 
\citep{Mamajek01}. Other young stars in the field toward the 4th Galactic quadrant, including 
the $\eta$ Cha cluster, TW Hya association, and $\beta$ Pic group, all have ages of 
$\sim$10 Myr and are moving radially away from LCC and UCL. Perhaps these 
stars formed in small molecular clouds that accumulated within the expanding LCC/UCL 
superbubble shells. 

\section{CONCLUSIONS}

From our spectroscopic survey of an X-ray- and kinematically-selected
sample of late-type stars in the Sco-Cen OB association, we summarize our
main findings as follows:

\begin{itemize}
\item We have identified a population of low-mass stars in the Lower Centaurus-Crux (LCC) 
and Upper Centaurus-Lupus (UCL) OB subgroups with the following properties:
(1) G-K spectral types, (2) subgiant surface gravities, (3) Lithium-rich, 
(4) strong X-ray emission (L$_X$ $\simeq$ 10$^{30}$ - 10$^{31}$ erg\,s$^{-1}$), 
and (5) proper motions consistent with the high-mass members. 
We classify stars which show these characteristics as bona fide pre-MS stars or
``post-T Tauri'' stars. X-ray and kinematic selection
(the RASS-ACT/TRC sample) yielded a hit rate of 93\% for selecting
probable pre-MS stars, while kinematic selection alone (dZ99
{\it Hipparcos} sample G-K dwarfs and subgiants) yielded 73\%.

\item We estimate the mean age of the pre-MS population in the LCC subgroup
to be 17\,$\pm$\,1\,Myr (DM97 tracks), 21\,$\pm$\,2\,Myr (PS01), and 
23\,$\pm$\,2\,Myr (SDF00). For UCL, the pre-MS population's
mean age is 15\,$\pm$\,1\,Myr (DM97), 19\,$\pm$\,1\,Myr (PS01), and 22\,$\pm$\,1\,Myr 
(SDF00). The UCL pre-MS estimate appears to be slightly biased towards younger ages 
(by $\sim$1 Myr) through contamination by Lupus or US members. We also calculate 
new MS turn-off ages of 16\,$\pm$\,1 Myr for LCC and 17\,$\pm$\,1 Myr for UCL using 
the dZ99 {\it Hipparcos} membership and \citet{Bertelli94} evolutionary tracks. 
The UCL pre-MS and turn-off age estimates are roughly self-consistent, and similar to
previously published estimates. Our age estimates for LCC (pre-MS and turn-off) 
are older than previous estimates, and are equal to or slightly older than UCL.

\item We find that 68\% of the low-mass star-formation in each subgroup took 
place within a $<$4-6 Myr span, and 95\% took place within $<$8-12 Myr 
(using DM97 tracks). The conditions were
right for producing low-mass stars in the LCC and UCL progenitor molecular clouds
for $<$10 Myr.

\item We find the frequency of CTTSs among a pre-MS population in an OB association 
with masses of 1.3\,$\pm$\,0.2 (1$\sigma$)\,\msol\, and ages of 
13\,$\pm$\,1\,(s.e.)\,$\pm$\,6\,(1$\sigma$) Myr (DM97 tracks) to be only 
0.9$^{+2.1}_{-0.8}$\% (1/110). 
The younger age results from using our entire (i.e. magnitude-biased) sample
of pre-MS stars. Only one star in our sample showed both strong H$\alpha$-emission and a K-band 
excess: the previously known CTTS TYC 9246-971-1 (\#34 = PDS 66 = Hen 3-892). 
This suggests that the disk accretion phase lasts $\leq$10-20 Myr in the evolution
of solar-type stars in OB associations.

\item We demonstrate that a surface gravity indicator for classifying field 
G and K stars (\ion{Sr}{2} $\lambda$4077 to \ion{Fe}{1} $\lambda$4071) 
can be used to distinguish whether Li-rich stars are pre-MS or ZAMS in nature. 
When this indicator is used in tandem with other youth diagnostics (Li abundance, 
X-ray emission, H$\alpha$ emission, and kinematics), one can confidently
classify a star as pre-MS in nature. 

\end{itemize}

\acknowledgments

EM and MM acknowledge support through NASA contract 1224768 administered by
JPL. EM and JL were also supported through a NASA/JPL grant. 
EM also thanks the $\Sigma\Xi$ Research Honor Society for a Grant-in-Aid 
which financed part of our observing run. We thank the TAC of the Mt. Stromlo and Siding Springs 
Observatories 2.3-metre telescope for their generous allocation of telescope time. 
This project benefited from discussions and help from J. Beiging, A. Blaauw, J. Carpenter,
C. Corbally, J. de\,Bruijne, T. de\,Zeeuw, M. Hamuy, L. Hillenbrand, W. Lawson, D. Soderblom, 
J. Stauffer, and S. Strom. We offer special thanks to Ronnie Hoogerwerf for assistance
in the target selection effort and discussions. 
This publication makes use of data products from the Two Micron All Sky Survey
(2MASS), which is a joint project of the Univ. of Massachusetts and the 
Infrared Processing and Analysis Center (IPAC), funded by NASA and NSF.

\appendix

\section{The Pre-Main Sequence \teff\, Scale \label{appendix_teff}}

Pre-main sequence stars lie between dwarfs (V) and subgiants (IV) on 
color-absolute magnitude or temperature-luminosity H-R diagrams. 
A given visual spectral type will 
correspond to cooler temperatures as surface gravity
decreases \citep[e.g.][]{Gray91,deJager87}. 
Dwarf temperature scales are often adopted for pre-MS populations, however
it is prudent to account for the effects of surface gravity.

We quantify the effects of \logg\, on the SpT vs. \teff\, relation
using two datasets. First, we fit a polynomial surface to \teff(SpT, \logg) using the
data from \citet[][Table 2]{Gray91}. As with most compilations
of \teff(SpT) in the FGK-star regime, we find that a trinomial
is the best low-order fit, and that a linear dependence on \logg\,
adequately accounts for the effects of surface gravity on temperature.
Our second method finds a similar
surface fit to \teff(SpT, \logg\,) using published \teff\, and \logg\, 
values \citep{Cayrel01} for the GK standards of \citet{Keenan89} and 
F standards of \citet{Garcia89}
(those within 0.3 dex of solar [Fe/H], and luminosity class IV and V only).
We adopt the isochrones from \citet{D'Antona97} for a fiducial
\logg\, value as a function of \teff\, for a coeval 15-Myr-old population.

Both assessments yield essentially the same result: dwarf temperature
scales for G-K stars should be lowered by 35\,K for a 15 Myr-old population. 
The temperature decrement increases for younger ages: 70\,K-40\,K for G0-K2 10-Myr-old stars,
235\,K-105\,K for G0-K2 5-Myr-old stars, and 260\,K-180\,K for G3-K2 1-Myr-old stars.
Both techniques yielded a linear dependence of \logg\,
on \teff(SpT, \logg\,), and the slopes were similar: 
$\partial$\teff/$\partial$\logg\, $\simeq$ 220\,K/log(cm s$^{-1}$)
for the Keenan standards with Cayrel de Strobel stellar atmosphere data, 
and $\partial$\teff/$\partial$\logg\, $\simeq$ 190 K/log(cm s$^{-1}$)
for the interpolation of Gray's (1991) Table 2. As the evolutionary model isochrones
are parallel to the main sequence when the stars are on the radiative tracks
(i.e. $\sim$10-30 Myr for $\sim$1\,\msol\, stars), the $\Delta$\logg\, between
a 15 Myr-isochrone and the main sequence is fairly constant over the
G-K spectral types. Hence one naively expects a linear offset
in \teff(SpT, \logg\,) between the 15-Myr isochrone and the MS. 

With the 1$\sigma$ scatter between published \teff(SpT)
relations being $\sim$60\,K amongst G stars, the systematic shift is nearly negligible. 
Upon comparing several temperature scales from the literature, we adopt
the dwarf \teff\, scale from \cite{Schmidt-Kaler82}, and apply a -35\,K offset to 
correct for the effects of lower surface gravity for a putative 
15-Myr-old population. We conclude that adopting dwarf \teff\, vs. SpT scales for pre-MS stars younger
than $\sim$10 Myr will systematically overestimate their \teff\,values,
and in turn, their masses inferred from evolutionary tracks. This could
have deleterious systematic effects on derived initial mass functions for 
young associations.

\section{Standards with Questionable Luminosity Class}

Several of the standard stars we observed had H-R diagram positions, published
\logg\, values, and \ion{Sr}{2} $\lambda$4077/\ion{Fe}{1} $\lambda$4071 ratios (SrFe) 
which differed from what is expected for their luminosity classes given in 
\citet{Keenan89}. The differences are only at the half of a luminosity-class level. 
We adopt the Keenan temperature types for all of his standard stars, however 
we revised the luminosity classes of these stars to bring their H-R diagram positions, SrFe 
index, and published \logg\, estimates into harmony (Table \ref{tab_stan_new}). 
The SrFe indices for the vast majority of the standards formed loci according to 
luminosity class (Fig. \ref{fig:srfe}), so we are comfortable using the index as an 
additional descriminent. The dwarf regression line in the gravity indicator vs. 
temperature indicator plot (Fig. \ref{fig:srfe}) was constructed using only
Keenan standards for which his luminosity classification agreed with published 
\logg\, values and the H-R diagram position. 

\placetable{tab_stan_new}

\section{Polynomial Fits \label{appendix_poly}}

$\bullet$ {\it MI6 vs. Spectral Type}: This flux ratio is Index 6 of
\citet{Malyuto97}. We measure the index in magnitudes (MI6 = 
-2.5log(f($\lambda\lambda$5125-5245)/f($\lambda\lambda$5245-5290))), and find the following
relation for Keenan and Garcia F0-K6 III-IV standards (Table \ref{tab_stan}) 
within 0.3 dex of solar metallicity:

\begin{equation}
SpT = 33.26\pm0.07 + (22.75\pm0.48)\times\,{MI6}~~(0.06 < {MI6} < 0.26) 
\end{equation}
\begin{equation}
SpT = 30.82\pm0.16 + (105.38\pm8.28)\times\,{MI6} - (711.74\pm173.81)\times\,{MI6}^2~~(-0.04 < {MI6} < 0.06)
\end{equation}

\noindent where SpT is the spectral type on Keenan's (1984) scale, 
i.e. F5 = 28, F8 = 29, G0 = 30, G2 = 31, G5 = 32, G8 = 33, K0 = 34, K1 = 35, and 
K2 = 36. Intermediate types can be assigned e.g. G9 = 33.5, G9.5 = 33.75, K0+ = 34.25, 
K0.5 = 34.5, etc. The first equation applies to K0-K6 stars, and the second equation applies to 
F0-K0 stars. The 1$\sigma$ dispersion in these fits is 0.6 subtypes. 

$\bullet$ {\it $\lambda$4374/$\lambda$4383 vs. Spectral Type}: 
This band ratio consists of two 3\AA\, bands centered on 4374.5\AA\, and 4383.6\AA. 
We measure the index in magnitudes as YFe = -2.5$\times$log(f($\lambda$4374.5)/f($\lambda$4383.6)).
We find the following relation between the band ratio and SpT for F0-K6 III-V 
stars:

\begin{equation}
SpT = 25.97\pm0.30 + (20.47\pm0.90)\,\times\,YFe
\end{equation}

\noindent The residual standard deviation to the fit (using 20 Keenan F0-K5 III-V standards
within 0.3 dex of solar metallicity) is 0.6 subtypes.

$\bullet$ {\it Surface Gravity Index \ion{Fe}{1} $\lambda$4071/\ion{Sr}{2} $\lambda$4077
vs. Spectral Type}: We measure a surface gravity index using the flux ratio of two 3\AA\, 
bands centered on \ion{Fe}{1} $\lambda$4071.4 and \ion{Sr}{2} $\lambda$4076.9. 
We measure the flux ratio in
magnitudes: SrFe = -2.5$\times$log(f($\lambda$4071)/f($\lambda$4077)), and plot
against our MI6 spectral type index. The Keenan standard dwarfs confirmed as being 
main sequence stars define a narrow locus:

\begin{equation}
SrFe = -0.078\pm0.005 + (2.123\pm0.261)\times\,MI6 - (8.393\pm2.945)\times\,MI6^2\,
+ (15.487\pm8.672)\times\,MI6^3
\end{equation}

\begin{figure}
\epsscale{1.0}
\plotone{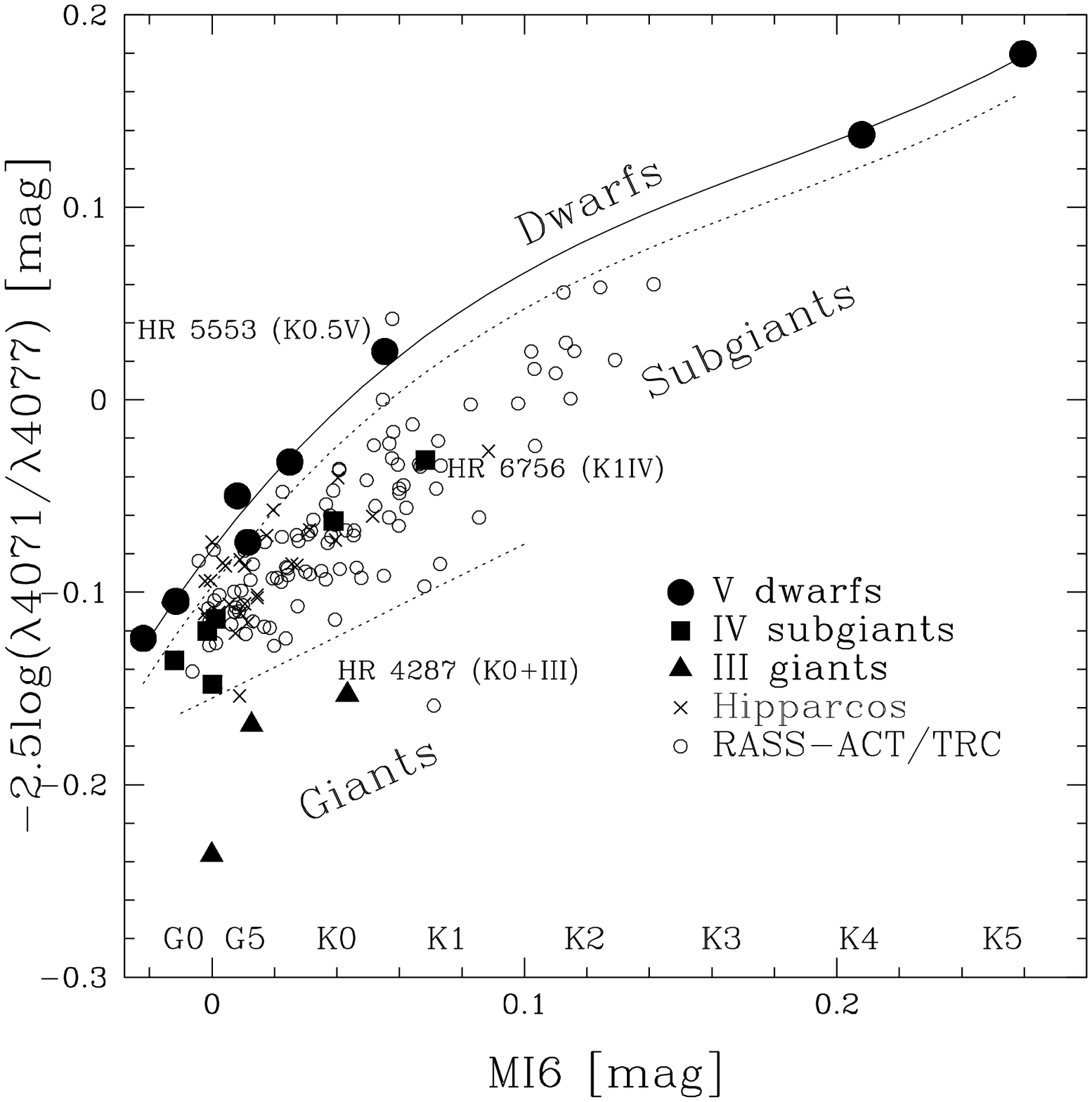}
\caption{ 
\label{fig:srfe}
The MI6 band-ratio (\teff\, indicator) vs. the band-ratio of \ion{Fe}{1} $\lambda$4071/
\ion{Sr}{2} $\lambda$4077 (surface gravity indicator). The solid line is a polynomial fit
to only the dwarf standards. The dashed lines separate dwarfs, subgiants, and giants.
The dwarf-subgiant dashed line is -2$\times$ the $\sigma$-residual below the dwarf 
regression, whereas the subgiant-giant boundary is placed somewhat arbitrarily to 
resolve the observed subgiant and giant loci. Empirically, this diagram suggests
that most of the target stars are consistent with being G and K-type subgiants, with
few giant and dwarf interlopers. A few early-K standards are noted for 
reference.
}

\end{figure}

\noindent The 1$\sigma$ sample standard deviation of this fit is 0.0094 mag in SrFe. The 
boundary between dwarfs and subgiants in Fig. \ref{fig:srfe} is -2$\sigma$ of the dwarf
locus. This relation is valid for F9-K6 stars. 

\begin{figure}
\epsscale{1.0}
\plotone{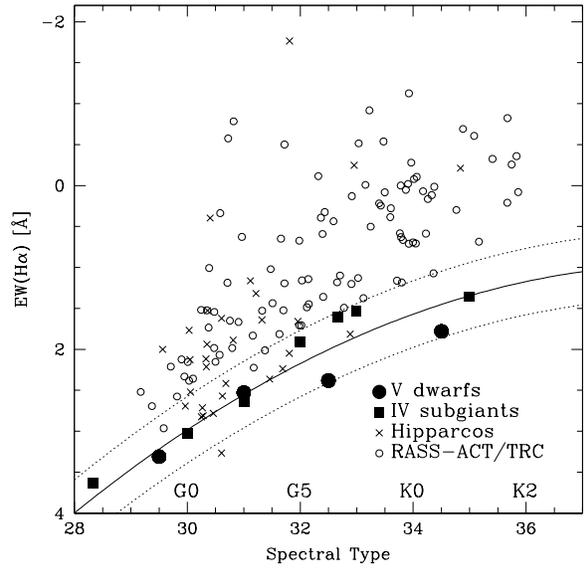}
\caption{ 
\label{fig:ha}
H$\alpha$ EWs for the pre-MS candidates compared to inactive field
dwarfs and subgiants. Symbols are the same as for Fig. \ref{fig:srfe}. 
The solid line is the average EW(H$\alpha$) for dwarf and subgiant standard stars. 
The dashed line represents the $\pm$2$\sigma$ residual scatter in the relation 
(encompassing all of the standards). Stars above this line
are clearly chromospherically active, however those within the 2$\sigma$
scatter have H$\alpha$ emission similar to older field stars. 
}
\end{figure}

\begin{figure}
\epsscale{1.0}
\plotone{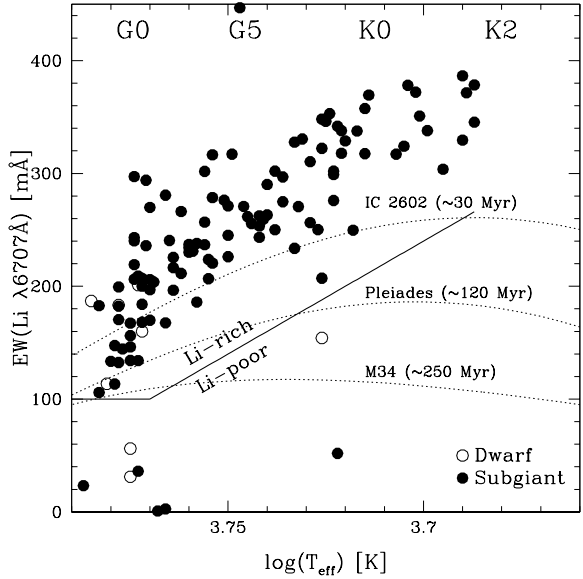}
\caption{ 
\label{fig:li}
EWs for \ion{Li}{1} $\lambda$6707 for the program stars compared
to regression fits for stars in young open clusters (see \S \ref{lithium}).
We discuss assignment of dwarf and subgiant luminosity classes in \S\ref{quantspt}.
The pre-MS candidates form an obvious locus, and we select all stars above the 
solid line as ``Li-rich''.
}
\end{figure}

\begin{figure}
\epsscale{1.0}
\plotone{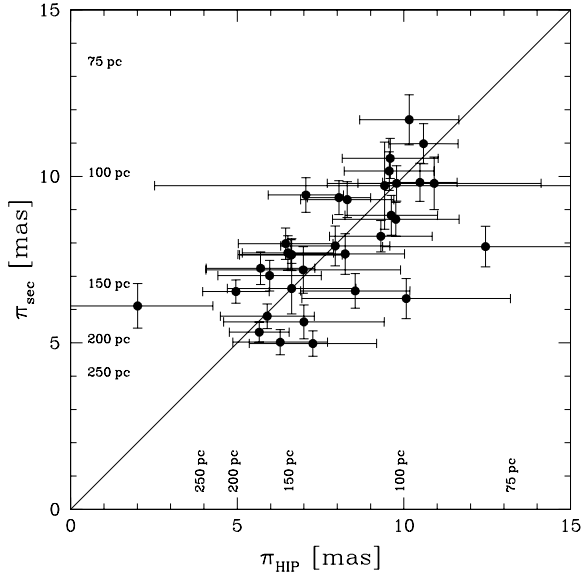}
\caption{
\label{fig:dist}
Comparison between {\it Hipparcos} astrometric parallaxes
and our secular parallaxes calculated using the moving group method.
Data points are Pre-MS (and Pre-MS?) association members from
Tables \ref{tab_rasstyc_pms} and \ref{tab_hip_pms}.
}
\end{figure}

\begin{figure}
\epsscale{1.0}
\plotone{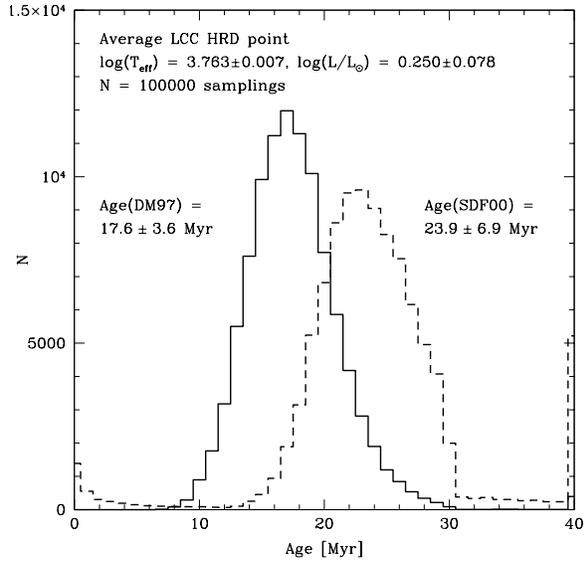}
\caption{ 
\label{fig:hist2}
A histogram of the inferred ages from the DM97 and SDF00 tracks for
a hypothetical LCC Pre-MS star with average H-R diagram point (\teff,\logl) and 
gaussian uncertainties. The extreme right bin retains all points older than 40 Myr. 
The standard deviations are calculated using only stars with ages between 1-100 Myr. 
}
\end{figure}

\begin{figure}
\epsscale{1.0}
\plotone{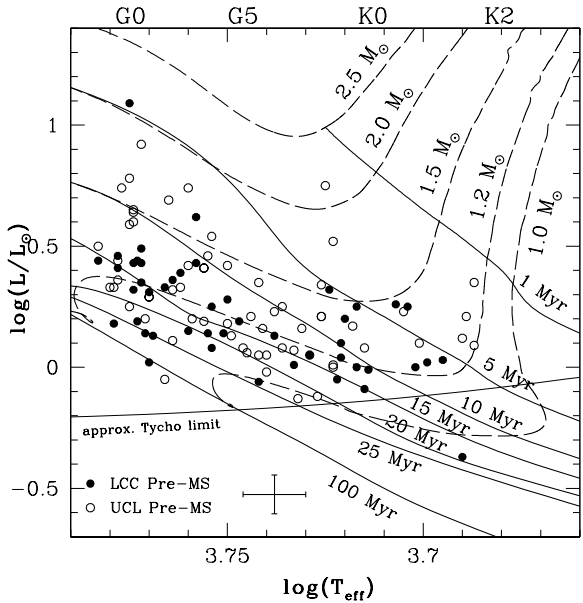}
\caption{
\label{fig:hrd}
Theoretical H-R diagram for stars identified as Pre-MS or Pre-MS? in Tables
\ref{tab_hip_pms} and \ref{tab_rasstyc_pms} in the UCL (open circles)
and LCC samples (filled circles). The pre-MS evolutionary tracks of DM97 are
overlayed. The ACT/TRC magnitude limit (V = 11 mag) is shown for a distance of 
150 pc (A$_V$ = 0.3 assumed). The star in the bottom right corner (TYC 8648-446-1) is one
of the faintest stars in our sample (V = 11.2 mag) with larger than average
errors in \logl\, - hence its unusual position. The average 1$\sigma$ 
error bars in \logt\, and \logl\, are shown.        
}
\end{figure}

\begin{figure}
\epsscale{1.0}
\plotone{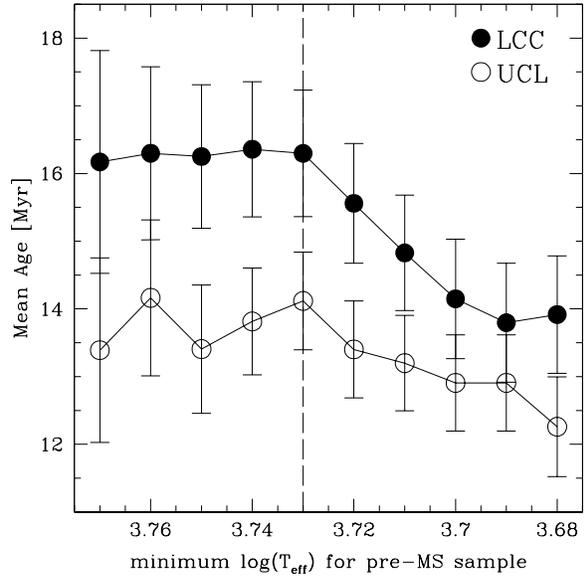}
\caption{\label{fig:agestat}
Illustration of the effects of magnitude bias on our mean age estimates for 
the pre-MS populations. The abscissa is the minimum log(\teff) threshold 
for evaluating the mean sample ages (using DM97 tracks). The ordinate is calculated
mean age with standard errors of the mean (shown; typically $\approx$1 Myr). 
At cooler temperatures (later than K0), the magnitude limit of our survey biases 
the sample towards more luminous stars, thereby decreasing the mean age estimate.
From this diagram, we choose \logt\,=\,3.73 (vertical dashed line) as the lower \teff\, 
cut-off for evaluating the mean pre-MS ages. 
Known spectroscopic binaries are included here, but excluded in the final 
age estimates presented in Table 8. The {\it observed} isochronal ages and 
spread are 16\,$\pm$\,5 Myr for LCC and 14\,$\pm$5 Myr for UCL.}
\end{figure}

\begin{figure}
\epsscale{1.0}
\plotone{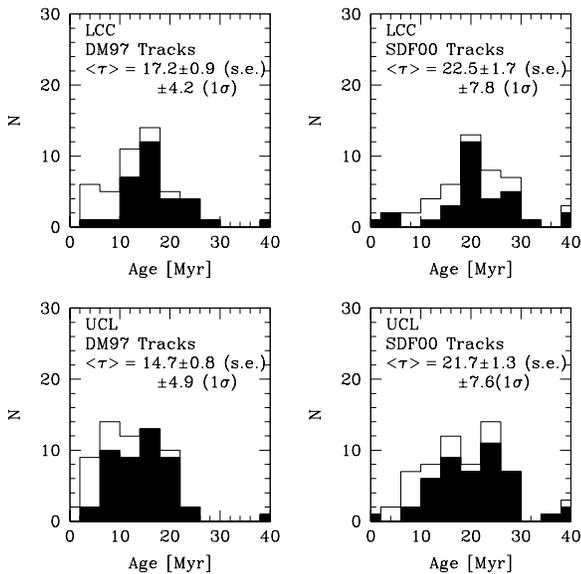}
\caption{
\label{fig:hist}
Histograms of the isochronal ages for pre-MS and ``pre-MS?'' candidates
in Tables \ref{tab_hip_pms} and \ref{tab_rasstyc_pms} from the
models of DM97 and SDF00. The filled bins are for stars with \logt\, $>$ 3.73, and the 
unfilled bins are for the entire (magnitude-biased) sample. Mean isochronal ages
(with standard errors of the means and 1$\sigma$ uncertainties) are given for the 
unbiased sample (\logt\, $>$ 3.73). Outliers with isochronal ages of $>$40 Myr are 
counted within the 40 Myr bin.
}
\end{figure}

\begin{figure}
\epsscale{1.0}
\plotone{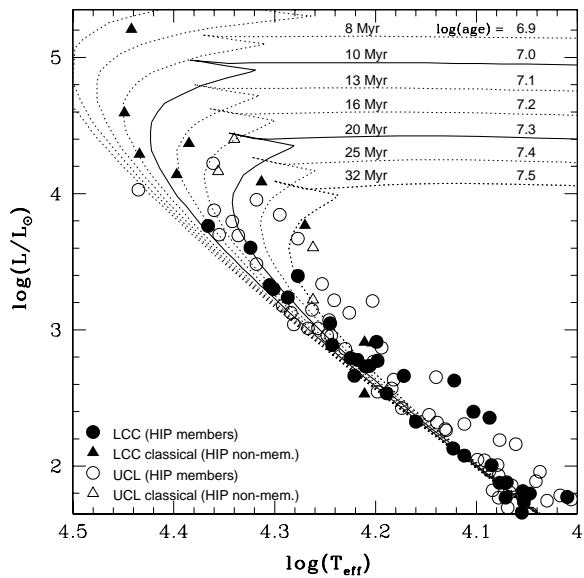}
\caption{
\label{fig:postms}
Theoretical H-R diagram for the B-star candidate members of the LCC \& UCL memberships 
using the evolutionary tracks of \citet{Bertelli94}. Only the most 
massive {\it Hipparcos} members were included in the age estimates.
The unusual variable HIP 67472 ($\mu$ Cen; B2Vnpe; \logl, \logt\, = 4.43, 4.0)
was excluded from the UCL turn-off age estimate.  
}
\end{figure}

\begin{figure}
\epsscale{1.0}
\plotone{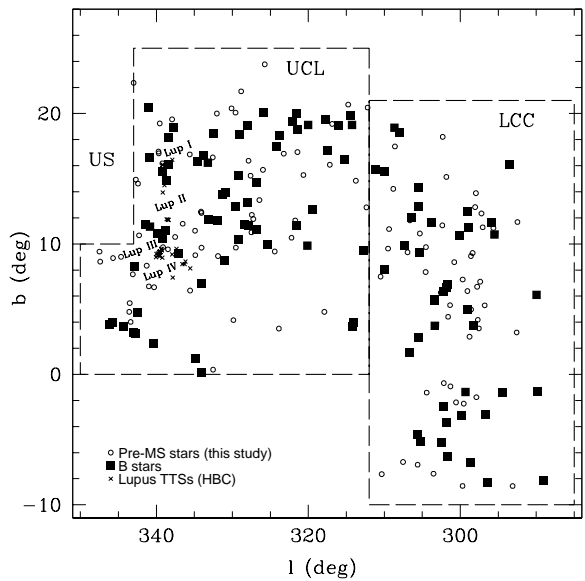}
\caption{
\label{fig:map}
Map of the UCL and LCC subgroups of the Sco-Cen OB association (Sco OB2). The
B-star population from \citet{deZeeuw99} is shown by filled squares. The pre-MS
(and pre-MS?) sample from this survey is shown as open circles. Pre-{\it ROSAT} 
T Tauri stars in the HBC catalog associated with the Lupus cloud are shown as 
Xs \citep{Herbig88}. 
}
\end{figure}

$\bullet$ {\it H$\alpha$ vs. Spectral Type}: In Fig. \ref{fig:ha}, we fit the 
equivalent widths of the H$\alpha$ feature (at low resolution, the photospheric
absorption plus the chromospheric emission) as a function of spectral type for 
F0-K6 dwarf and subgiant standard stars (Table \ref{tab_stan}) 
with the following polynomial:

\begin{equation}
EW(H\alpha) = 2.983\pm0.066 - (0.456\pm0.027)\times\,(SpT\,-\,30) + (2.574\pm0.378)\times10^{-2}\times\,(SpT\,-\,30)^2
\end{equation}

\noindent EW(H$\alpha$) is measured in \AA. SpT is spectral type on Keenan's scale 
(as before). The sample standard deviation of the polynomial fit to 11 standards was 0.20\AA. 

$\bullet$ {\it Converting Tycho $\bvj$ to Cousins-Johnson $\bv$}: 
The {\it Hipparcos} catalog gives linear relations
between $\bvt$, $\bvj$, $V$, and $V_T$ for stars of a wide range in spectral types.
\citet{Bessell00} compared the {\it Hipparcos}/Tycho photometry and that of the E-region
photometric standards, and refined the relations between the two systems. Table 2 of 
\citet{Bessell00} gives a standard relation between $\bvt$, Cousins-Johnson $\bvj$, and
(V - V$_T$) for B-G dwarfs and K-M giants. We fit the following relations to Bessell's tables:

\begin{equation}
V = V_T + 9.7\times10^{-4} - 1.334\times10^{-1}(\bvt) + 
5.486\times10^{-2}(\bvt)^2 -\\
1.998\times10^{-2}(\bvt)^3
\label{tycho1}
\end{equation}
\begin{equation}
\bv = (\bvt) + 7.813\times10^{-3}(\bvt) - 1.489\times10^{-1}(\bvt)^2 + 3.384\times10^{-2}(\bvt)^3
\label{tycho2}
\end{equation}
\begin{equation}
\bv  = (\bvt)  - 0.006 - 1.069\times10^{-1}(\bvt) + 1.459\times10^{-1}(\bvt)^2
\label{tycho3}
\end{equation}

The V($V_T$, $\bvt$) polynomial equation \ref{tycho1} applies to stars from 
-0.25 $<$ $(\bvt)$ $<$ 2.0 (B-M types). Equation \ref{tycho2} is for stars with 
0.5 $< (\bvt) <$ 2.0, and equation \ref{tycho3} is for stars with 
-0.25 $< (\bvt) <$ 0.5. We do not quote uncertainties in the polynomial
coefficients since the Bessell relations are already smoothed. 
These equations fit Bessell's standard relations to 1-2 millimagnitudes.





\clearpage




\clearpage
\begin{deluxetable}{ccccccccccl}
\tabletypesize{\scriptsize}
\setlength{\tabcolsep}{0.01in} 
\rotate
\tablewidth{0pt}
\tablecaption{Properties of de Zeeuw {\it et al.} (1999) Sco-Cen G-K Candidate Members\label{tab_hip}}
\tablehead{
{(1)}  &{(2)}   &{(3)}                                 & {(4)}                          &{(5)}& {(6)}   & {(7)} & {(8)}  & {(9)} & {(10)} & {(11)}\\
{Name} &{OB}    & {$\alpha$~~$\delta$ (J2000.0)}       & {$\mu_{\alpha*}$, $\mu_\delta$}& {V} & {$\bvj$}& {SpT} & {J}    & {H}   & {K$_s$}& {notes} \\
{HIP}  &{Grp.}  & {h~~m~~s~~$^\circ$~~\arcmin~~\arcsec}& {(mas/yr)}                     &{(mag)}& {(mag)}   & {MSS} & {(mag)}  & {(mag)} & {(mag)}  & 
}
\startdata
57524& LCC& 11:47:24.55 -49:53:03.0& -33.7$\pm$0.8  -10.2$\pm$1.0 & 9.07 & 0.63$\pm$0.02 & G3/5VPq2    & 7.91$\pm$0.02 & 7.59$\pm$0.06 & 7.51$\pm$0.02 &1RXS J114724.3-495250, TWA 19A\\ 
58996& LCC& 12:05:47.48 -51:00:12.1& -37.5$\pm$0.8  -11.5$\pm$0.8 & 8.89 & 0.63$\pm$0.02 & G1Vq1       & 7.67$\pm$0.01 & 7.37$\pm$0.03 & 7.27$\pm$0.01 &var(0.06), 1RXS J120547.8-510007\\     
59854& LCC& 12:16:27.84 -50:08:35.8& -29.1$\pm$1.3  ~-8.6$\pm$1.0 & 9.34 & 0.67$\pm$0.02 & G3Vq1       & 8.04$\pm$0.01 & 7.69$\pm$0.02 & 7.61$\pm$0.03 &1RXS J121627.9-500829\\
60885& LCC& 12:28:40.05 -55:27:19.3& -34.9$\pm$0.8  -16.0$\pm$0.7 & 8.89 & 0.64$\pm$0.02 & G1q4        & 7.68$\pm$0.02 & 7.40$\pm$0.05 & 7.28$\pm$0.03 &var(0.02), 1RXS J122840.3-552707\\    
60913& LCC& 12:29:02.25 -64:55:00.6& -37.5$\pm$1.2  -11.4$\pm$0.9 & 9.04 & 0.73$\pm$0.02 & G5Vq1       & 7.60$\pm$0.01 & 7.26$\pm$0.01 & 7.14$\pm$0.01 &...\\
62445& LCC& 12:47:51.87 -51:26:38.2& -30.4$\pm$1.0  ~-8.7$\pm$1.0 & 9.52 & 0.80$\pm$0.03 & G8/K0Vq3    & 7.79$\pm$0.01 & 7.38$\pm$0.03 & 7.25$\pm$0.01 &var(0.10), 1RXS J124751.7-512638\\   
63797& LCC& 13:04:30.96 -65:55:18.5& -42.4$\pm$1.0  -14.2$\pm$0.8 & 8.48 & 0.74$\pm$0.01 & G3Vq1       & 7.02$\pm$0.04 & 6.79$\pm$0.02 & 6.71$\pm$0.01 &... \\
63847& LCC& 13:05:05.29 -64:13:55.3& -36.5$\pm$1.1  -19.5$\pm$1.3 & 9.18 & 0.73$\pm$0.02 & G5Vq4       & 7.78$\pm$0.01 & 7.44$\pm$0.03 & 7.36$\pm$0.03 &var(0.03)      \\
\enddata
\tablecomments{
Columns: 
(1) {\it Hipparcos} ID, 
(2) OB subgroup region, 
(3) J2000.0 position, 
(4) proper motion components (mas yr$^{-1}$; where $\mu_{\alpha*} = \mu_{\alpha}$cos$\delta$), 
(5) V magnitude (Johnson), 
(6) $\bvj$ color (Johnson), 
(7) Spectral types from the Michigan Spectral Survey (Vol. 1-3); ``q1'' indicates the flag 
``quality = 1'' in the MSS catalog, where q=1,2 stars are judged to have reliable spectral 
types, 
(8-10) 2MASS JHK$_s$ magnitudes, 
(11) Notes on variability, X-ray counterparts, and multiple star system name (CCDM) from 
{\it Hipparcos}.  ``{\it Var.(N.NN)}'' indicates that the H6 field in the {\it Hipparcos} 
catalog is identified as being variable in the broad $H_p$ pass-band. The magnitude scatter 
{\it N.NN} in $H_p$ is listed, and period {\it P} if found. 
Near-IR photometry is from the preliminary 2MASS database. The name is given
of RASS-BSC X-ray sources within 40'' of {\it Hipparcos} stars; many of the 
variable stars are also {\it ROSAT} sources. Astrometric data and optical photometry 
are from from the {\it Hipparcos} catalog \citet{ESA97}.
}
\end{deluxetable}


\clearpage
\begin{deluxetable}{lccccccccccccl}
\tabletypesize{\scriptsize} 
\setlength{\tabcolsep}{0.02in} 
\rotate
\tablewidth{0pt}
\tablecaption{Properties of \rasstyc\, Candidates \label{tab_rasstyc}}
\tablehead{
{(1)}      & {(2)}& {(3)}                     & {(4)}                        & {(5)}   & {(6)}    & {(7)}& {(8)} & {(9)}  & {(10)} & {(11)}   & {(12)}  & {(13)}   & {(14)}\\ 
{Name}     &{OB}  & {$\alpha,\delta$(J2000)}  & {$\mu_{\alpha*},~~~\mu_\delta$} & {V$_T$} & {$\bvt$} & {J}  & {H}   & {K$_s$}& {1RXS} & {sep.}   & {X-ray} & {Hrdns.} & {Table 5}\\
{TYC}      &{Grp.} & {h~~m~~s~~$^\circ$~~\arcmin~~\arcsec} & {(mas yr$^{-1}$)} & {(mag)}   & {(mag)}    & {(mag)}& {(mag)} & {(mag)}  & {Name} & {(\arcsec)}& {(ct s$^{-1}$)} & {HR1} & {\#}} 
\startdata
9212-2011-1 &LCC & 10:57:49.38 -69:13:59.9 & -36.1$\pm$2.4  ~~5.7$\pm$2.3 & 10.49$\pm$0.04 & 1.09$\pm$0.08 & 8.49$\pm$0.03 & 8.02$\pm$0.05 & 7.79$\pm$0.01 & J105751.2-691402 & 10 & 1.61e-01 & 0.31  &1  \\
8625-388-1  &LCC & 11:32:08.34 -58:03:20.0 & -41.6$\pm$3.1  ~-1.4$\pm$3.0 & 10.02$\pm$0.03 & 1.02$\pm$0.06 & 8.09$\pm$0.02 & 7.64$\pm$0.04 & 7.48$\pm$0.02 & J113209.3-580319 &  7 & 8.25e-02 & 0.03  &2  \\
8222-105-1  &LCC & 11:35:03.76 -48:50:22.0 & -24.7$\pm$2.0  ~-7.0$\pm$1.9 & 10.35$\pm$0.03 & 0.75$\pm$0.05 & 9.22$\pm$0.02 & 8.93$\pm$0.03 & 8.86$\pm$0.03 & J113501.5-485011 & 24 & 1.01e-01 & 0.49  &... \\
8982-3046-1 &LCC & 12:04:14.42 -64:18:51.7 & -28.9$\pm$3.9  ~-4.7$\pm$3.7 & 10.09$\pm$0.03 & 0.75$\pm$0.05 & 8.82$\pm$0.01 & 8.50$\pm$0.02 & 8.40$\pm$0.02 & J120413.3-641837 & 16 & 2.77e-01 & 0.32  &... \\
8982-3213-1 &LCC & 12:04:48.87 -64:09:55.4 & -32.9$\pm$3.5  ~+0.8$\pm$3.1 &  9.49$\pm$0.02 & 0.79$\pm$0.03 & 8.05$\pm$0.01 & 7.71$\pm$0.01 & 7.61$\pm$0.01 & J120448.2-640942 & 13 & 3.47e-01 & 0.46  &3  \\ 
8640-2515-1 &LCC & 12:06:13.54 -57:02:16.8 & -24.6$\pm$2.8  ~-7.7$\pm$2.6 & 10.77$\pm$0.05 & 0.81$\pm$0.08 & 9.25$\pm$0.03 & 8.84$\pm$0.03 & 8.77$\pm$0.03 & J120613.9-570215 &  3 & 1.44e-01 & 0.27  &4  \\ 
8644-802-1  &LCC & 12:09:41.86 -58:54:45.0 & -33.8$\pm$3.1  -11.8$\pm$2.8 & 10.21$\pm$0.03 & 1.26$\pm$0.07 & 8.31$\pm$0.01 & 7.81$\pm$0.01 & 7.66$\pm$0.01 & J120941.5-585440 &  5 & 2.33e-01 & 0.16  &5  \\ 
8644-340-1  &LCC & 12:11:31.43 -58:16:53.2 & -36.0$\pm$2.6  ~-8.7$\pm$1.8 & 10.29$\pm$0.04 & 1.00$\pm$0.07 & 8.58$\pm$0.01 & 8.09$\pm$0.03 & 7.96$\pm$0.02 & J121131.9-581651 &  4 & 2.41e-01 & 0.09  &6  \\
\enddata
\tablecomments{
Columns: 
(1) Tycho-2 name, 
(2) OB subgroup region, 
(3) J2000.0 position from Tycho-2 catalog, 
(4) proper motion ($\alpha$*, $\delta$ components) in mas yr$^{-1}$ (where $\mu_{\alpha*} = \mu_{\alpha}$cos$\delta$), 
(5) V$_T$ magnitude, 
(6) $(\bv)_{T}$ color, 
(7-9) 2MASS JHK$_s$ magnitudes, 
(10) X-ray counterpart name in RASS-BSC, 
(11) optical-X-ray separation, 
(12) X-ray count rate (ct s$^{-1}$), 
(13) Hardness ratio HR1,
(14) Number in Table 5 if star is found to be ``pre-MS'' or ``pre-MS?'' in nature. 
X-ray data is from {\it ROSAT} All-Sky Survey BSC \citep{Voges99}, astrometry and optical photometry are from {\it Tycho-2} 
catalog \citep{Hog00TYC2a}, and near-IR photometry is from the 2MASS working database.
}
\end{deluxetable}


\clearpage
\begin{deluxetable}{rcccrccclll}
\tablewidth{0pt}
\tabletypesize{\scriptsize} 
\rotate
\tablecaption{Spectral Standard Stars \label{tab_stan}}
\tablehead{
{Name} &{Name} & {MK}  & {V}   & {$\pi\,\pm\,\sigma_\pi$} & {$\bvj$} & {[Fe/H]}  & {MSS}     & {Adopted} &{notes} \\
{HD}   &{HR}   & {Sp.Type} & {(mag)} & {(mas)}                  & {(mag)}     & {adopted} & {Sp.Type} & {Sp.Type} & {...}}
\startdata
182640 & 7377 & F0IV              & 3.36 & 65.1\,$\pm$\,0.8 & 0.32 &    ...\, & F2V       & F0IV     & \tablenotemark{a,b}\\
173677 & 7061 & F6V               & 4.19 & 52.4\,$\pm$\,0.7 & 0.48 &  --0.1\, & ...       & F6IV-V   & \tablenotemark{c}\\
 84117 & 3862 & F9V               & 4.93 & 67.2\,$\pm$\,0.7 & 0.53 &    ...\, & G0V       & F9V      & \\
121370 & 5235 & G0IV              & 2.68 & 88.2\,$\pm$\,0.8 & 0.58 &    0.2\, & ...       & G0IV     & \tablenotemark{b}\\
 89010 & 4030 & G1.5IV-V          & 5.95 & 32.9\,$\pm$\,0.9 & 0.66 &    0.0\, & ...       & G1.5IV-V & \\
126868 & 5409 & G2IV              & 4.84 & 24.2\,$\pm$\,1.0 & 0.69 &    0.0\, & G3V       & G2III-IV & \tablenotemark{a,d}\\
161239 & 6608 & G2IIIb            & 5.73 & 26.1\,$\pm$\,0.6 & 0.68 &    ...\, & ...       & G2IV     & \\
146233 & 6060 & G2Va              & 5.49 & 71.3\,$\pm$\,0.9 & 0.65 &    0.0\, & G5V       & G2V      & \tablenotemark{a}\\
\enddata
\tablecomments{
SB1,2 = spectroscopic binary in either SIMBAD or Duquennoy \& Mayor (1991). MSS = Michigan Spectral 
Survey Vols. 1-5 = \citet{Houk75}, \citet{Houk78}, \citet{Houk82}, \citet{Houk88}, \citet{Houk99}. 
The [Fe/H] estimate is adopted from the compilation of published values in \citet{Cayrel01}.
A semi-colon after the [Fe/H] value indicates considerable scatter ($>$0.2 dex) in the published estimates.}
\tablenotetext{a}{SIMBAD lists as variable or suspected var., however {\it Hipparcos} finds scatter in Hp $\leq$ 
0.015 mag. The typical scatter for the other standard stars was 0.005 mag in Hp ({Hipparcos} magnitude), 
with none greater than 0.01 mag.}
\tablenotetext{b}{Spectroscopic binary}
\tablenotetext{c}{From standard list of \citet{Garcia89}, and originally in \citet{Johnson53}, but not listed in 
either \citet{Keenan88} or \citet{Keenan89}.} 
\tablenotetext{d}{HR 5409 is a resolved binary (sep. 5'') listed by SIMBAD as a variable star, however {\it Hipparcos} found the 
scatter in the Hp band to be only 0.015 mag.}
\tablenotetext{e}{HR 5072 is the planet host 70 Vir. \citet{Keenan89} call it G4V, however it is G5V in virtually
every other reference (e.g. \citet{Gray01}). We retain Keenan's classification.} 
\tablenotetext{f}{{\it Hipparcos} ``G'' binary}
\end{deluxetable}


 \clearpage
\begin{deluxetable}{cccccl}
\tablewidth{0pt}
\tabletypesize{\scriptsize} 
\setlength{\tabcolsep}{0.01in} 
\tablecaption{Stellar Classification Scheme \label{tab_class}}
\tablehead{
{Li-}   & {H$\alpha$} & {Lum.}  & {N(R-T)} & {N(HIP)} & {Adopted}  \\
{Rich}  & {Excess}    & {Class} & {\#}     & {\#}     & {Class.}   
}
\startdata
Yes & Yes     & IV    & 94  & 7 &  Pre-main sequence (PMS) \\
Yes & No      & IV    &  3  & 6 &  Probably pre-main sequence? (PMS?) \\
Yes & Strong  & IV    &  1  & 0 &  Pre-MS Classical T Tauri star (CTTS)\tablenotemark{a} \\
Yes & Yes/No  & V     &  4  & 1 &  Young Dwarf (ZAMS) \\
No  & Yes     & V     &  1  & 0 &  Active Dwarf \\
No  & Yes     & IV    &  4  & 0 &  Active Subgiant \\
No  & No      & IV    &  1  & 2 &  Subgiant \\
No  & No      & III   &  0  & 2 &  Giant \\
No  & No      & V     &  0  & 2 &  Dwarf (MS)\\
No  & No(Wide)& IV    &  2  & 0 &  Chromosph.-Active Binary (CAB)\tablenotemark{b}\\
\tableline
... & ...     & ...   & 107 & 20 & Total \\
\enddata
\tablecomments{N(R-T) is number in \rasstyc\, sample.
N(HIP) is number in {\it Hipparcos} sample.
``Li-rich'' implies significant Li absorption, and
above the line in Fig.\ref{fig:li}. ``Active'' means that the H$\alpha$
equivalent widths are $>$2 $\times$ the $\sigma$-residual above the regression 
of values for field, standard stars (Fig. \ref{fig:ha}; Appendix \ref{appendix_poly}), 
implying that chromospheric emission is filling in the absorption
line. Luminosity classes are assigned according to a star's placement
in Fig. \ref{fig:srfe}.
}
\tablenotetext{a}{included in Pre-MS sample count.}
\tablenotetext{b}{included in subgiant sample count.}
\end{deluxetable}


\clearpage
\begin{deluxetable}{cccccccccccccccrrrrl}
\renewcommand{\arraystretch}{.5}
\tablewidth{0pt}
\tabletypesize{\tiny} 
\rotate
\tablecaption{LCC \& UCL Pre-Main Sequence Members from \citet{deZeeuw99} List \label{tab_hip_pms}}
\tablehead{
{(1)}   & {(2)} & {(3)}                     & {(4)}        &{(5)}        & {(6)}  & {(7)}  & {(8)}  & {(9)}       & {(10)}  & {(11)}  & {(12)}       & {(13)} & {(14)}        & {(15)}       & {(16)}     & {(17)}     & {(18)}     & {(19)} & {(20)}\\ 
{Name}  & {OB}  & {$\alpha,\delta$(J2000)}  & {$\pi_{HIP}$}&{$\pi_{sec}$}& {P$_1$}& {P$_3$}& {Spec.}& {log(\teff)}& {\logl} & {A$_V$} & {EW(H$\alpha$)} & {EW(Li)}   & {log(L$_X$)}  & {log(L$_X$/} & {DM97}     & {PS01}     & {SDF00}     & {Class}& {Notes}\\
{HIP \#}& {Grp} & {h m s d \arcmin \arcsec} & {(mas)}      &{(mas)}      & {\%}   & {\%}   & {Type} & {(K)}       & {(dex)} & {(mag)} & {(\AA)}& {(\AA)} & {(erg s$^{-1}$)}& {L$_{bol}$)} & {age,mass} & {age,mass} & {age,mass} & {...}  & {...}}
\startdata
57524 & LCC & 11:47:24.55 -49:53:03.0 &  9.62$\pm$1.39 &  8.83$\pm$0.60 &  81 & 97 & F9IV    & 3.783$\pm$0.004 & 0.44$\pm$0.06 & 0.30$\pm$0.10 & 2.0 & 0.18 &   30.8 &   -3.3 & 16 1.3 &  16 1.2 & 20 1.3 & PMS  & HD 102458, TWA 19\\  
58996 & LCC & 12:05:47.48 -51:00:12.1 &  9.78$\pm$1.16 &  9.79$\pm$0.52 &  91 & 99 & G1IV    & 3.772$\pm$0.005 & 0.43$\pm$0.05 & 0.14$\pm$0.04 & 3.3 & 0.21 &   30.5 &   -3.6 & 14 1.3 &  13 1.3 & 18 1.3 & PMS? & HD 105070\\ 
59854 & LCC & 12:16:27.84 -50:08:35.8 &  8.24$\pm$1.78 &  7.67$\pm$0.61 &  99 &100 & G1IV    & 3.772$\pm$0.006 & 0.49$\pm$0.07 & 0.37$\pm$0.11 & 1.6 & 0.20 &   30.9 &   -3.2 & 13 1.4 &  12 1.3 & 17 1.3 & PMS  & HD 106725\tablenotemark{a}\\  
60885 & LCC & 12:28:40.05 -55:27:19.3 &  7.06$\pm$1.13 &  9.49$\pm$0.52 &  14 & 78 & G0IV    & 3.778$\pm$0.005 & 0.46$\pm$0.05 & 0.16$\pm$0.05 & 2.7 & 0.13 &   30.3 &   -3.8 & 14 1.3 &  13 1.3 & 18 1.3 & PMS? & HD 108568\\ 
60913 & LCC & 12:29:02.25 -64:55:00.6 & 10.48$\pm$1.12 &  9.82$\pm$0.58 & 100 &100 & G4.5IV  & 3.758$\pm$0.008 & 0.43$\pm$0.05 & 0.21$\pm$0.04 & 2.0 & 0.19 &$<$29.7 &$<$-4.3 & 11 1.4 &  12 1.3 & 16 1.3 & PMS? & HD 108611\\ 
62445 & LCC & 12:47:51.87 -51:26:38.1 &  6.61$\pm$1.54 &  7.70$\pm$0.55 &  13 & 75 & G4.5IVe & 3.758$\pm$0.013 & 0.62$\pm$0.06 & 0.66$\pm$0.06 &-1.8 & 0.24 &   30.7 &   -3.4 &  7 1.6 &   5 1.6 & 11 1.5 & PMS  & SB3?, V940 Cen\\ 
63847 & LCC & 13:05:05.29 -64:13:55.3 &  9.56$\pm$1.35 & 10.18$\pm$0.61 &  48 & 89 & G3IV    & 3.764$\pm$0.007 & 0.36$\pm$0.06 & 0.33$\pm$0.06 & 1.6 & 0.22 &$<$29.7 &$<$-4.3 & 14 1.3 &  15 1.2 & 19 1.3 & PMS  & HD 113466\\ 
65517 & LCC & 13:25:47.83 -48:14:57.9 &  9.59$\pm$1.44 & 10.56$\pm$0.57 &  50 & 91 & G1.5IV  & 3.770$\pm$0.006 & 0.02$\pm$0.05 & 0.14$\pm$0.08 & 1.9 & 0.17 &   30.4 &   -3.3 & .. 1.0 &  39 1.0 &... 1.0 & PMS  & V966 Cen\tablenotemark{b}\\
\enddata
\end{deluxetable}


\clearpage
\begin{deluxetable}{llcccccccccccccrrrll}
\tablewidth{0pt}
\renewcommand{\arraystretch}{.5}
\tabletypesize{\tiny}
\rotate
\tablecaption{Pre-Main Sequence LCC \& UCL \rasstyc\, Members \label{tab_rasstyc_pms}}
\tablehead{
{(1)}& {(2)}     &{(3)} & {(4)}                     & {(5)}        & {(6)}  & {(7)}  & {(8)}  & {(9)}       & {(10)}  & {(11)}  & {(12)}       & {(13)} & {(14)}        & {(15)}       & {(16)}     & {(17)}     & {(18)}     & {(19)}  & {(20)}  \\ 
{ID} & {Name}    &{OB}  & {$\alpha,\delta$(J2000)}  & {$\pi_{sec}$}& {P$_1$}& {P$_3$}& {Spec.}& {log(\teff)}& {\logl} & {A$_V$} & {EW(H$\alpha$)}         & {EW(Li)}   & {log(L$_X$)}  & {log(L$_X$/} & {DM97}     & {PS01}     & {SDF00}    & {Class} & {Names}   \\
{\#} & {TYC}     &{Grp} & {h m s d \arcmin \arcsec} & {(mas)}      & {\%}   & {\%}   & {Type} & {(K)}       & {(dex)} & {(mag)} & {(\AA)}& {(\AA)} & {(erg s$^{-1}$)}& {L$_{bol}$)} & {age,mass} & {age,mass} & {age,mass} & {...}   & {\& Notes} 
}
\startdata
1  & 9212-2011-1 & LCC & 10:57:49.38 -69:13:59.9 &  9.76$\pm$1.01 &  79 &  94 &  K1+IV    & 3.699$\pm$0.010 & 0.02$\pm$0.09 & 0.35$\pm$0.06 & ~ 0.7 & 0.34 & 30.3 & -3.3 &  7 1.2 &  13 1.1&  15 1.2&  PMS  & \\
2  & 8625-388-1  & LCC & 11:32:08.34 -58:03:20.0 & 10.78$\pm$1.21 &  59 &  84 &  G7IV     & 3.747$\pm$0.013 & 0.19$\pm$0.10 & 0.68$\pm$0.13 & ~ 0.4 & 0.45 & 29.9 & -3.7 & 15 1.2 &  19 1.1&  23 1.2&  PMS  & HD 304428; SB2\\  
3  & 8982-3213-1 & LCC & 12:04:48.87 -64:09:55.4 &  8.36$\pm$1.27 &  12 &  49 &  G1IV     & 3.773$\pm$0.005 & 0.44$\pm$0.13 & 0.60$\pm$0.11 & ~ 1.5 & 0.21 & 30.8 & -3.2 & 14 1.3 &  13 1.3&  18 1.3&  PMS  & HD 104919\\  
4  & 8640-2515-1 & LCC & 12:06:13.54 -57:02:16.8 &  6.48$\pm$1.06 &  90 &  97 &  G4IV     & 3.760$\pm$0.007 & 0.15$\pm$0.14 & 1.45$\pm$0.06 & ~ 0.6 & 0.23 & 30.6 & -3.1 & 21 1.1 &  25 1.1&  28 1.2&  PMS  & \\  
5  & 8644-802-1  & LCC & 12:09:41.86 -58:54:45.0 &  9.00$\pm$1.14 &  65 &  88 &  K0IVe    & 3.720$\pm$0.013 & 0.20$\pm$0.11 & 0.28$\pm$0.10 & -0.1  & 0.33 & 30.5 & -3.2 &  7 1.4 &  12 1.2&  15 1.3&  PMS  & \\  
6  & 8644-340-1  & LCC & 12:11:31.43 -58:16:53.2 &  9.29$\pm$0.91 &  95 &  99 &  G9IV     & 3.729$\pm$0.008 & 0.05$\pm$0.09 & 0.28$\pm$0.10 & ~ 0.3 & 0.31 & 30.5 & -3.2 & 14 1.2 &  21 1.1&  24 1.1&  PMS  & PPM 770135\\  
7  & 9231-1566-1 & LCC & 12:11:38.14 -71:10:36.0 & 10.15$\pm$0.68 &  91 &  98 &  G3.5IV   & 3.762$\pm$0.005 & 0.39$\pm$0.06 & 0.48$\pm$0.08 & ~ 1.4 & 0.27 & 30.5 & -3.5 & 13 1.3 &  13 1.3&  18 1.3&  PMS  & HD 105923\\  
8  & 8636-2515-1 & LCC & 12:12:35.75 -55:20:27.3 &  9.24$\pm$0.90 &  50 &  84 &  K0+IV    & 3.717$\pm$0.007 & 0.00$\pm$0.09 & 0.48$\pm$0.12 & ~ 0.2 & 0.34 & 30.3 & -3.2 & 11 1.2 &  19 1.1&  23 1.1&  PMS  & PPM 770145\\
\enddata
\end{deluxetable}



\clearpage
\begin{deluxetable}{lcccccccl}
\tablewidth{0pt}
\tabletypesize{\scriptsize} 
\rotate
\tablecaption{\rasstyc\, \& {\it Hipparcos} Stars Rejected as Sco-Cen Members \label{tab_others}}
\tablehead{
{(1)}         &         {(2)}           & {(3)}  & {(4)}  & {(5)} & {(6)}        & {(7)} & {(8)}       & {(9)}\\
{Name}        &{$\alpha,\delta$(J2000)} & {P$_1$}& {P$_3$}& {SpT} & {EW(H$\alpha$)} & {EW(Li)}  & {log(L$_X$/}& {Type}\\
{}            &{h m s d \arcmin \arcsec}& {\%}   & {\%}   & {...} & {(\AA)}& {(\AA)}& {L$_{bol}$)}& {...} }
\startdata
TYC 8222-105-1  & 11:35:03.76 -48:50:22.0 &  71 & 93 & F8.5V    & 2.7 &0.19 & -3.4    & ZAMS \\
TYC 8982-3046-1 & 12:04:14.42 -64:18:51.7 &  97 & 99 & G1V      & 2.0 &0.20 & -3.1    & ZAMS \\
TYC 8990-701-1  & 13:13:28.11 -60:00:44.6 &   9 & 48 & F9V      & 2.2 &0.11 & -3.4    & ZAMS \\
TYC 8285-847-1  & 14:16:57.91 -49:56:42.3 &  43 & 87 & G2.5IV   & 2.2 &0.00 & -3.9    & subgiant (CAB)\\ 
TYC 7833-1106-1 & 15:08:00.55 -43:36:24.9 &   4 & 53 & G2IV     & 0.6 &0.00 & -3.5    & active subgiant\\ 
TYC 8293-92-1   & 15:09:27.93 -46:50:57.2 &  28 & 71 & K0IV     & 0.7 &0.05 & -3.4    & active subgiant\\ 
TYC 7318-593-1  & 15:31:21.93 -33:29:39.5 &  94 & 98 & G9.5V    & 0.0 &0.15 & -3.3    & active dwarf   \\ 
TYC 7858-526-1  & 16:38:38.47 -39:33:03.5 &  98 &100 & F8.5IV   & 2.5 &0.02 & -4.0    & active subgiant (CAB) \\
\enddata
\tablecomments{
Columns:
(1) Name from Tycho-2 or {\it Hipparcos} catalogs,
(2) J2000.0 position,
(3) Membership probability (with $v_{disp}$ = 1 \kms),
(4) Membership probability (with $v_{disp}$ = 3 \kms),
(5) Spectral type (see \S\ref{spt}),
(6) Equivalent width of H$\alpha$ $\lambda$6562.8 (\AA), 
(7) Corrected EW of \ion{Li}{1} $\lambda$6707.8 (\AA),
(8) Logarithm of ratio between X-ray and bolometric luminosities (approximate upper limits assume RASS 
PSPC detection limit of 0.05 ct s$^{-1}$ and HR1 = 0),
(9) Class of object
}

\tablenotetext{a}{HIP 65423 = HD 116402. \citet{Cutispoto02} measure EW(Li) = 220 m\AA, and $v$sin$i$ =
35 \kms.}

\end{deluxetable}


 \clearpage
\begin{deluxetable}{ccccc}
\tablewidth{0pt}
\tabletypesize{\scriptsize} 
\setlength{\tabcolsep}{0.01in} 
\tablecaption{Age Estimates of LCC \& UCL \label{tab_age}}
\tablehead{
{Ref.} & {Tracks} & {method} & {age(LCC)} & {age(UCL)} \\
{}     & {}       & {}        & {(Myr)}    & {(Myr)} 
}
\startdata
1   & 4    & pre-MS  & 17\,$\pm$\,1& 15\,$\pm$\,1\\
1   & 5    & pre-MS  & 21\,$\pm$\,2& 19\,$\pm$\,1\\
1   & 6    & pre-MS  & 23\,$\pm$\,2& 22\,$\pm$\,1\\
\tableline 
1   & 7    & turn-off& 16\,$\pm$\,1& 17\,$\pm$\,1\\ 
\tableline 
2   & 8    & turn-off& 11-12       & 14-15       \\
3   & 8,9  & turn-off& 10-11       & 12-13       \\
\tableline
\enddata
\tablecomments{Uncertainties are standard errors of
the mean. Pre-MS age estimates exclude known SBs and stars with \logt\,$<$ 3.73, which
bias the calculated ages. The turn-off ages from this work are determined using only 
early-B stars classified as members by \citet{deZeeuw99}.} 
\tablerefs{(1) this work, (2) \citet{deGeus89}, (3) \citet{deZeeuw85}, 
(4) \citet{D'Antona97}, (5) \citet{Palla01}, (6) \citet{Siess00},
(7) \citet{Bertelli94}, (8) \citet{Maeder81}, (9) Cogan (priv. comm.) 
}
\end{deluxetable}
	

 \clearpage
\begin{deluxetable}{lccccc}
\tablewidth{0pt}
\tabletypesize{\scriptsize} 
\setlength{\tabcolsep}{0.01in} 
\tablecaption{Revised Luminosity Classes of Standard Stars \label{tab_stan_new}}
\tablehead{
{Star}  & {Published}  & {\logg}     & {Sr/Fe} & {HRD } & {adopted}\\
{}      & {Lum. class} & {estimates} & {class} & {pos.} & {Lum. class} 
}
\startdata
HR 5072 & V            & 3.8-3.9 (IV-V)  & IV-V    & IV-V   & IV-V \\
HR 4995 & IV-V         & 3.0-3.7 (III/IV)& IV      & IV     & IV   \\ 
HR 5409 & IV           & 3.3-3.9 (III/IV)& III-IV  & III-IV & III-IV \\
HR 6608 & IIIb         & ...             & IV      & IV     & IV \\
\tableline
\enddata
\tablecomments{Published luminosity classes from \citet{Keenan89}.
The range of published \logg\, estimates come from the compilation of \citet{Cayrel01}.
``Sr/Fe class'' is from measuring the \ion{Fe}{1}\,$\lambda$4071/\ion{Sr}{2}\,$\lambda$4071 
ratio in our spectra and intercomparison to the spectral standards in Table \ref{tab_stan}
(see Fig. \ref{fig:srfe}).
The luminosity class from the HRD position uses the V and $\bvj$ data from {it Hipparcos}
and the standard relations from Appendix B of \citet{Gray91}.
}

\end{deluxetable}


\end{document}